\documentclass[english,aps,prd,superscriptaddress,preprintnumbers,nofootinbib,preprint]{revtex4-1}
\usepackage[T1]{fontenc}
\usepackage[utf8]{inputenc}
\usepackage{verbatim}

\usepackage{empheq}
\usepackage{amssymb}
\usepackage{amsmath}
\usepackage{amsfonts}

\usepackage[colorlinks]{hyperref}
\hypersetup{
 citecolor=blue,
 linkcolor=blue,
 urlcolor=blue}

\usepackage{dcolumn}
\usepackage{graphicx}
\usepackage{etoolbox} 

\newcommand{\as}{\alpha_s}

\newcommand{\MSbar}{\overline{\rm MS}}

\newcommand{\be}{\begin{equation}}
\newcommand{\ee}{\end{equation}}
\newcommand{\bea}{\begin{eqnarray}}
\newcommand{\eea}{\end{eqnarray}}
\newcommand{\bes}{\begin{equation}\begin{split}}

\newcommand{\LO}{{\rm LO}}
\newcommand{\NLO}{{\rm NLO}}
\newcommand{\NNLO}{{\rm NNLO}}

\newcommand{\fb}{{\rm fb}}

\newcommand{\gev}{{\rm GeV}}

\newcommand{\bb}{ {b\bar{b}} }

\newcommand{\kt}{k_t}

\newcommand{\ptv}{p_{t,V}}

\newcommand{\ptVcut}{\ptv > 150~\gev}


\usepackage{babel}
\usepackage{color}
\begin{document}
\global\long\def\order#1{\mathcal{O}\left(#1\right)}
\global\long\def\d{\mathrm{d}}
\global\long\def\P{P}
\global\long\def\amp{{\mathcal M}}
\preprint{TTP21-017}
\preprint{P3H-21-044}
\preprint{OUTP-21-15P}
\preprint{CERN-TH-2021-086}

\def\KIT{
  Institute for Theoretical Particle Physics, KIT, Karlsruhe, Germany
}
\def\IKP{
  Institute for Astroparticle Physics, KIT, Karlsruhe, Germany
}
\def\CERN{
  Theoretical Physics Department, CERN, 1211 Geneva 23, Switzerland
}
\def\OX{
  Rudolf Peierls Centre for Theoretical Physics, Clarendon Laboratory, Parks Road,
  Oxford OX1 3PU, UK and  Wadham College, Oxford OX1 3PN, UK
}

\title{ Anomalous couplings in associated $VH$ production with
    Higgs decay to massive $b$ quarks at NNLO in QCD}

\author{Wojciech~Bizo\'{n}}
\email[Electronic address: ]{wojciech.bizon@kit.edu}
\affiliation{\KIT}
\affiliation{\IKP}

\author{Fabrizio~Caola}
\email[Electronic address: ]{fabrizio.caola@physics.ox.ac.uk}
\affiliation{\OX}

\author{Kirill~Melnikov}
\email[Electronic address: ]{kirill.melnikov@kit.edu}
\affiliation{\KIT}

\author{Raoul~R\"ontsch }
\email[Electronic address: ]{raoul.rontsch@cern.ch}
\affiliation{\CERN}

\begin{abstract}
  We combine the NNLO QCD description of Higgs boson production in
  association with an electroweak vector boson $V = W~{\rm or}~Z$ with
  a similarly-precise description of Higgs boson decays into a pair of
  massive $b$ quarks and with the anomalous couplings that modify
  interactions of the Higgs and electroweak vector bosons.  The
  resulting numerical code provides the most advanced theoretical tool
  to investigate such anomalous couplings in the associated Higgs
  boson production process.  We study the impact of anomalous
  couplings on fiducial cross sections and differential distributions
  and argue that, with increased QCD precision, smaller anomalous
  couplings become accessible in kinematic regions where the effects
  of higher-dimensional operators in the Standard Model Effective
  Field Theory remain small and the EFT expansion is under control.

\end{abstract}

\maketitle

\section{Introduction}

Studies of Higgs boson properties in experiments at the Large Hadron
Collider (LHC) have converged to the conclusion \cite{couplATLCMS}
that the Standard Model of particle physics describes Higgs couplings
to gauge bosons and to (some) matter fields with a precision between
10 and 30 percent.  To reach a higher precision, new experimental
measurements as well as refined theoretical descriptions of major
Higgs production processes are needed. The forthcoming Run III of the
LHC, as well as its high-luminosity phase, will play an important role
in achieving these goals.

From a theoretical perspective, some of the simplest but also most
interesting Higgs boson production processes are those where Higgs
bosons are produced in association with vector bosons, i.e.  $pp \to
WH$ and $pp \to ZH$.  Indeed, at lowest order in perturbative QCD,
both of these processes are of the Drell-Yan type $pp \to V^* \to VH$,
so that their description through next-to-next-to-leading order (NNLO)
in perturbative QCD is quite straightforward.  These production
processes allow for a study of Higgs-gauge interactions.

Associated production processes also provide an environment in which
the decay of the Higgs to a $b \bar b$ pair can be
observed~\cite{Butterworth:2008iy,Aaboud:2017xsd,Sirunyan:2017elk,Aaboud:2018zhk,Sirunyan:2018kst},
allowing the study of the Higgs coupling to $b$ quarks.  Therefore,
describing both the production $pp \to VH$ and the decay $H \to b \bar
b$ processes with the same precision is important.  Moreover, it is
important to consider $b$ quarks as massive since in this case one can
apply conventional jet algorithms to identify $b$ jets and reconstruct
Higgs boson kinematics.  Indeed, it was shown recently in
Ref.~\cite{Behring:2020uzq} that working with massless $b$ quarks may
lead to sizeable differences in theoretical predictions for the
associated production process $pp \to WH$.

The status of theoretical predictions for $VH$ processes in the
Standard Model (SM) is quite advanced. Refined predictions that
include both
QCD~\cite{Han:1991ia,Baer:1992vx,Ohnemus:1992bd,Mrenna:1997wp,Spira:1997dg,
  Djouadi:1999ht,Brein:2003wg,Brein:2011vx,Brein:2012ne,Ferrera:2011bk,
  Ferrera:2013yga,Ferrera:2014lca,
  Campbell:2016jau,Ferrera:2017zex,Caola:2017xuq,Gauld:2019yng} and
electroweak~\cite{Ciccolini:2003jy,Denner:2011id,Denner:2014cla}
radiative corrections are available. Recently, NNLO QCD corrections
for $WH$ production in association with a hard jet were
computed~\cite{Gauld:2020ced}.

A major difference between the $WH$ and $ZH$ final states is that,
starting from NNLO QCD, the latter receives large contributions from
the $gg\to ZH$ process. The $\mathcal O(\alpha_s^2)$ contribution of
this process has been known for a long time~\cite{Kniehl:1990iva,
  Dicus:1988yh}. Approximate results for the $\mathcal O(\alpha_s^3)$
contributions due to $gg \to ZH$ suggest that these can be quite
large~\cite{Altenkamp:2012sx,Harlander:2014wda} and should be included
for reliable predictions despite being formally subleading. In the
recent past, significant effort went into their
computation~\cite{Hasselhuhn:2016rqt,
  Davies:2020drs,Chen:2020gae,Alasfar:2021ppe}, using either numerical
methods or phenomenologically-motivated analytic approximations.
Strategies to extract the $gg\to ZH$ contribution from experimental
data have been investigated in Ref.~\cite{Harlander:2018yns}.  Both
the $WH$~\cite{giuliaWH,Alioli:2019qzz} and $ZH$~\cite{giuliaZH,
  Alioli:2019qzz,Bizon:2019tfo} processes have been matched to parton
shower Monte Carlo tools retaining NNLO QCD
accuracy. Ref.~\cite{Bizon:2019tfo} also includes NNLO QCD corrections
to the $H\to b \bar b$ decay.  Dedicated parton shower Monte Carlo
tools including both next-to-leading order (NLO) QCD and electroweak
corrections~\cite{Granata:2017iod} and a refined treatment of the
$gg\to ZH$ contribution~\cite{Hespel:2015zea} also exist.

Similarly, advanced theoretical predictions for the $H\to b\bar b$
decay are available. The total rate is known up to N$^4$LO QCD in the
limit of massless bottom quarks~\cite{Gorishnii:1990zu,
  Gorishnii:1991zr, Kataev:1993be, Surguladze:1994gc, Larin:1995sq,
  Chetyrkin:1995pd, Chetyrkin:1996sr,
  Baikov:2005rw,Davies:2017xsp,Davies:2017xsp}. Electroweak
corrections are also known~\cite{Fleischer:1980ub, Bardin:1990zj,
  Dabelstein:1991ky, Kniehl:1991ze}. A comprehensive review of
computations of the $H\to b \bar b$ inclusive branching ratio and its
uncertainties can be found in
Refs.~\cite{Denner:2011mq,Spira:2016ztx}. At the differential level,
QCD corrections for massless $b$ quarks are known at
NNLO~\cite{Anastasiou:2011qx, DelDuca:2015zqa, Caola:2017xuq,
  Caola:2019pfz} and N$^3$LO~\cite{Mondini:2019vub,Mondini:2019gid}.
NNLO QCD results retaining the full $b$-quark mass dependence have
been presented in Refs.~\cite{Bernreuther:2018ynm,
  Behring:2019oci,Behring:2020uzq, Somogyi:2020mmk}.  Top-quark
effects have been studied in Refs.~\cite{Primo:2018zby,
  Mondini:2020uyy}.

Given this level of sophistication, it is interesting to extend the
precise modeling of associated Higgs boson production and $H \to b
\bar b$ decay to cases where Higgs couplings to gauge and matter
fields differ from the ones in the Standard Model Lagrangian.  A
convenient way to do this is provided by the Standard Model Effective
Field Theory (SMEFT), see e.g. Ref.~\cite{Brivio:2017vri} for a
review.  In principle, one may aim at the complete description of the
processes $pp \to VH(b \bar b)$ in the SMEFT, taking into account
contributions of {\it all} dimension-six operators present in the
SMEFT Lagrangian. However, since in this case the number of operators
that one has to consider becomes quite large, it makes sense to first
restrict oneself to a subset of operators. A particularly interesting
choice is those that modify the couplings of the Higgs boson to
electroweak gauge bosons. Indeed, this approach has been adopted in
Ref.~\cite{Mimasu:2015nqa} where this process was studied with NLO QCD
accuracy. The main goal of this work is to promote this analysis to
full NNLO QCD.

The paper is organized as follows. In Section~\ref{sec:sm} we briefly
describe the computation of NNLO QCD corrections to the processes $pp
\to VH(b\bar b)$, with $V = W,Z$, in the Standard Model.  Such a
computation for the $WH$ final state was discussed earlier in
Ref.~\cite{Behring:2020uzq}; the results for $ZH(b\bar b)$ with
massive $b$ quarks are new.  In Section~\ref{sec:anom} we describe
calculations that include both anomalous couplings and NNLO QCD
corrections to Higgs boson associated production and its decay into a
$b \bar b$ pair. We consider scenarios where anomalous couplings
modify fiducial cross sections only slightly so that the availability
of highly precise theoretical predictions for fiducial cross sections
and kinematic distributions becomes important. We conclude in
Section~\ref{sec:summ}.

 \section{Associated production $pp \to VH(b \bar b)$ in the Standard Model}
\label{sec:sm}

In this section, we briefly describe the computation of NNLO QCD
radiative corrections to the associated production process $pp \to
VH(b\bar b)$ in the Standard Model, keeping the $b$-quark masses
nonzero.  We note that the $WH(b\bar b)$ final state was discussed
recently in Ref.~\cite{Behring:2020uzq}. The results for the $ZH(b\bar
b)$ final state, which we mostly focus on in this section, are new.

 As we already mentioned in the introduction, the computation of NNLO QCD
 radiative corrections to $pp \to VH(b\bar b)$ involves two major
 ingredients: QCD corrections to the production process $pp \to VH$
 and QCD corrections to the decay process $H \to b \bar b$.
 An
 earlier computation of NNLO QCD corrections to $pp \to WH$ with the
 decay $H \to b \bar b$ for massless $b$ quarks was described in
 Ref.~\cite{Caola:2017xuq}.
 This computation was based on the nested
 soft-collinear subtraction scheme introduced in
 Ref.~\cite{Caola:2017dug}, and employed simple analytic formulas derived
 for the production and decay processes of color-singlet states in
 Refs~\cite{Caola:2019nzf, Caola:2019pfz}.

 This earlier computation was recently extended by including Higgs
 boson decays to {\it massive} $b$ quarks \cite{Behring:2020uzq},
 using predictions for $H \to b \bar b$ decay from
 Ref.~\cite{Behring:2019oci} and modifying the calculation of NNLO QCD
 corrections to the production process in Ref.~\cite{Caola:2017xuq} to
 exclude $b$ quarks from being active partons in a proton. As we
 already mentioned in the introduction, working with massive quarks
 allows us to employ conventional jet algorithms to describe
 $b$-flavored jets.

 In this paper, we have extended the above computations to the $ZH$
 final state.  From the point of view of soft and collinear
 subtractions, such an extension is straightforward since the analytic
 formulas derived in Ref.~\cite{Caola:2019nzf} are applicable to all
 color-singlet final states. For this reason, a transition from the
 $WH$ final state to the $ZH$ final state only requires us to change
 the relevant matrix elements and adjust flavors of colliding partons.
 However, an important difference between $ZH$ and $WH$ final states
 is the contribution of the $gg \to ZH$ partonic process which only
 exists in the former case.  Thanks to a large gluon flux, this
 contribution is significant; accounting for it in the theoretical
 prediction for the $ZH$ final state is important, especially for
 large values of $ZH$ invariant masses. In our calculation, we have
 included the exact $\mathcal O(\alpha_s^2)$ contributions to the
 $gg\to ZH$ channel but we have neglected higher-order terms which so
 far are not available.  Apart from corrections to the $gg$ channel,
 there are other classes of contributions proportional to the top
 Yukawa coupling for which the exact two-loop amplitudes are
 unknown. Here we followed the approach of
 Ref.~\cite{Campbell:2016jau}, which is based on the analysis of
 Ref.~\cite{Brein:2011vx}. In the notation of
 Ref.~\cite{Brein:2011vx}, we have included the so-called $V_{I,II}$
 contributions keeping only the leading term in the $m_{\rm
   top}\to\infty$ expansion.  We have also included exact $R_{I}$
 contributions, but discarded $R_{II}$ terms since they have been
 shown to be very small for phenomenologically relevant
 setups~\cite{Brein:2011vx}. Similarly, we have neglected effects of
 top quark loops in Drell-Yan type diagrams $pp\to V^*, V^*\to VH$ as
 they too have negligible phenomenological impact~\cite{Brein:2011vx}.

 As the first step in our discussion, we present cross sections and
 differential distributions for the two associated production processes
\begin{align}
  p p &\to W^+ H \to (\nu_e e^+) (\bb )\,, \\
  p p &\to Z H~ \to (e^- e^+) (\bb )\,,
\end{align}
at the $13~{\rm TeV}$~LHC.
We treat both decay processes $V \to \ell \bar\ell$ and $ H \to \bb$
in the narrow-width approximation.
Following Ref.~\cite{Behring:2020uzq},
we write differential cross sections
as
\be
   {\rm d} \sigma_{pp \to VH(b \bar b)}
     \propto {\rm Br}(H \to b \bar b)  \times {\rm d} \sigma_{pp \to VH} \times
     \frac{{\rm d} \Gamma_{H \to b \bar b} }{\Gamma_{H \to b \bar b}},
     \label{eq:sigExp}
\ee
and we do not perform an expansion of ${\rm Br}(H \to b \bar b)$ in a
series of $\alpha_s$, treating it as an input parameter.

For numerical computations we take ${\rm Br}(H \to b \bar b ) =
0.5824$ as recommended by the Higgs Cross Section Working
Group~\cite{hxswg}.  We set the Higgs boson mass to $M_H = 125~\gev$,
the vector boson masses to $ M_W = 80.399~\gev$ and $M_Z
=91.1876~\gev$, respectively, the {\it on-shell} $b$-quark mass to
$m_b= 4.78~\gev$, and the top quark mass to $m_t = 173.2~\gev$.
We use the Fermi
constant $G_F = 1.16639 \times 10^{-5}~\gev^{-2}$ and the weak mixing angle $\sin^2\theta_W = 0.2226459$.
The widths of
vector bosons are taken to be $\Gamma_W = 2.1054~\gev$ and
$\Gamma_Z = 2.4952~\gev$. Finally, we take the CKM matrix to be diagonal.

\begin{table}
  \centering
  \begin{tabular}{lccc}
    \hline
    Order
    & $\sigma^{W^+H}_{\rm fid}~[\fb]$
    & $\sigma^{ZH}_{\rm fid}~[\fb]$
    & $\sigma^{ZH+ggZH}_{\rm fid}~[\fb]$
    \\
    \hline
    $\LO$
    & $  21.22^{ +0.76}_{ -0.95}$
    & $  5.13^{ +0.17}_{ -0.21}$
    & $-$
    \\
    \hline
    $\NLO$
    & $  24.48^{ +0.42}_{ -0.25}$
    & $  5.87^{ +0.10}_{ -0.06}$
    & $-$
    \\
    \hline
    $\NNLO$
    & $  23.86^{ +0.16}_{ -0.17}$
    & $  5.69^{ +0.03}_{ -0.03}$
    & $  6.52^{ +0.25}_{ -0.19}$
    \\
    \hline
  \end{tabular}
  \caption{Fiducial cross sections for $pp \to W^+H \to (\nu_e e^+)
    (\bb )$ and $pp \to ZH \to (e^-e^+)(\bb)$ at the $13~{\rm TeV}$
    LHC at various orders of QCD perturbation theory calculated with
    massive $b$ quarks.  We set the factorization and renormalization
    scales equal to each other, $\mu_r = \mu_f = \mu$.  We use $\mu=
    \tfrac{1}{2} \sqrt{(p_V+p_H)^2}$ for the central value and the
    uncertainties are calculated by varying the scale $\mu$ by a
    factor of two in both directions.
    See text for details.}
  \label{tab:fid-xsec}
\end{table}

We note that the $b$-quark Yukawa coupling that enters the $H \to \bb$
decay rate is computed using the $\MSbar$ $b$-quark mass calculated at
$\mu = M_H$,
$\overline{m}_b(\mu=M_H)=2.81~\gev$~\cite{Chetyrkin:2000yt,Herren:2017osy}.
However,
since the physical cross sections in Eq.~\eqref{eq:sigExp} are
proportional to the ratio ${\rm d} \Gamma_{H \to b \bar b} / \Gamma_{H
  \to b \bar b}{}$, the dependence on the Yukawa coupling to a large
extent cancels out in the results that are presented below.

We define $W^+H$ and $ZH$ final states using kinematic selection
criteria for charged leptons and $b$-flavored jets.  To this end, we
require that an event contains at least two $b$ jets that are defined
with the anti-$\kt$ jet algorithm
\cite{Cacciari:2008gp,Cacciari:2011ma} and we choose the jet radius
$R=0.4$. Pseudo-rapidities and transverse momenta of charged leptons
and $b$ jets should satisfy the following constraints
\begin{align}
  \begin{aligned}
    &|\eta_{l}|   < 2.5 \,, \quad p_{t,l} > 25~\gev\,,\\
    &|\eta_{j_b}| < 2.5 \,, \quad p_{t,j_b} > 20~\gev\,.
  \end{aligned}
  \label{eq:cuts}
\end{align}
 Finally, following experimental analyses, we may 
additionally require that the vector boson has large transverse
momentum, $\ptVcut$. We always state explicitly when this cut is applied.

The fiducial cross sections are calculated with the four-flavor parton
distribution function (PDF) set \verb+NNPDF31_nnlo_as_0118_nf_4+. We emphasize that we employ
NNLO PDFs to compute LO, NLO and NNLO cross sections in what follows.
Moreover, we use $\as(M_Z) = 0.118$ and perform the running of the
strong coupling at three loops with five active flavors.

For all numerical results presented in this paper, the central value
of the renormalization and factorization scales {\it in the production
  process} is set to one half of the invariant mass of the $VH$
system, i.e.  $\mu_r = \mu_f = \mu = \tfrac{1}{2} \sqrt{(p_V+p_H)^2}$.  The
renormalization scale {\it for the decay process} is set to the Higgs
boson mass, $\mu_{r,\rm{dec}} = M_H$; it is kept fixed for all results reported in this paper.
The uncertainty of the cross sections is obtained by varying the scale
in the {\it production process}  by a factor of two around the central value.

\begin{table}
  \centering
  \begin{tabular}{lccc}
    \hline
    Order
    & $\sigma^{W^+H}_{\rm fid}~[\fb]$
    & $\sigma^{ZH}_{\rm fid}~[\fb]$
    & $\sigma^{ZH+ggZH}_{\rm fid}~[\fb]$
    \\
    \hline
    $\LO$
    & $ 3.89$
    & $ 0.97$
    & $-$
    \\
    \hline
    $\NLO$
    & $ 4.79^{ +0.13}_{ -0.10}$
    & $ 1.20^{ +0.03}_{ -0.03}$
    & $-$
    \\
    \hline
    $\NNLO$
    & $ 4.79^{ +0.02}_{ -0.06}$
    & $ 1.22^{ +0.03}_{ -0.03}$
    & $ 1.52^{ +0.11}_{ -0.09}$
    \\
    \hline
  \end{tabular}
  \caption{Fiducial cross sections in the boosted region ($\ptVcut$) for $pp \to W^+H \to (\nu_e e^+) (\bb )$ and
    $pp \to ZH \to (e^-e^+)(\bb)$ at the $13~{\rm TeV}$ LHC at various orders of
    QCD perturbation theory calculated with massive $b$ quarks.
    We set the factorization and renormalization scales equal to each other,  $\mu_r = \mu_f = \mu$.
    We use $\mu=  \tfrac{1}{2} \sqrt{(p_V+p_H)^2}$ for the central value and the uncertainties are calculated by varying the scale $\mu$ by a factor of two in both directions. 
    See text for details.}
  \label{tab:fid-xsec-boost}
\end{table}

We present fiducial cross sections for the above set of cuts at
leading order (LO), next-to-leading order (NLO) and NNLO in QCD for
the $W^+H$ and $ZH$ production processes in Tables~\ref{tab:fid-xsec}
and \ref{tab:fid-xsec-boost}.  The contribution of the gluon-initiated
process $gg \to ZH$ is reported separately. As we already mentioned,
this contribution is rather large.  Indeed, it follows from
Tables~\ref{tab:fid-xsec} and \ref{tab:fid-xsec-boost} that it
increases the fiducial cross section by 15 percent if no cut on the
$Z$ transverse momentum is applied, and by about 25 percent if the $Z$
boson is required to have a transverse momentum in excess of $150~{\rm
  GeV}$.\footnote{We note that in the $gg\to ZH$ channel there is a
  strong cancellation between the box and triangle diagrams, which
  however is only active for SM couplings. If e.g. the top Yukawa
  coupling were to be different, a very strong enhancement of this
  contribution could be expected~\cite{Englert:2013vua}. For example,
  if the top Yukawa coupling had the opposite
  sign~\cite{Farina:2012xp,Sirunyan:2018lzm}, the NNLO cross section
  in Table~\ref{tab:fid-xsec} would increase to about $10~\fb$.}

We also see from Tables~\ref{tab:fid-xsec} and \ref{tab:fid-xsec-boost} that  uncertainties of  {\it quark-initiated}
contributions at NNLO are less than a percent if no $p_{t,V}$ cut is applied, and around two percent in the presence of this cut.
The inclusion of the $gg \to ZH$ contribution increases
the uncertainty significantly, to about four percent without the additional $p_{t,V}$ cut and to about seven percent if we require 
 $\ptVcut$.
To reduce this uncertainty, the $gg \to ZH$ contribution
has to be computed at NLO in perturbative QCD and, as we already mentioned in the introduction,
a significant effort in this direction is currently underway~\cite{Hasselhuhn:2016rqt,
  Davies:2020drs,Chen:2020gae,Alasfar:2021ppe}.

\begin{figure}\centering
  \includegraphics[width=0.49\textwidth]{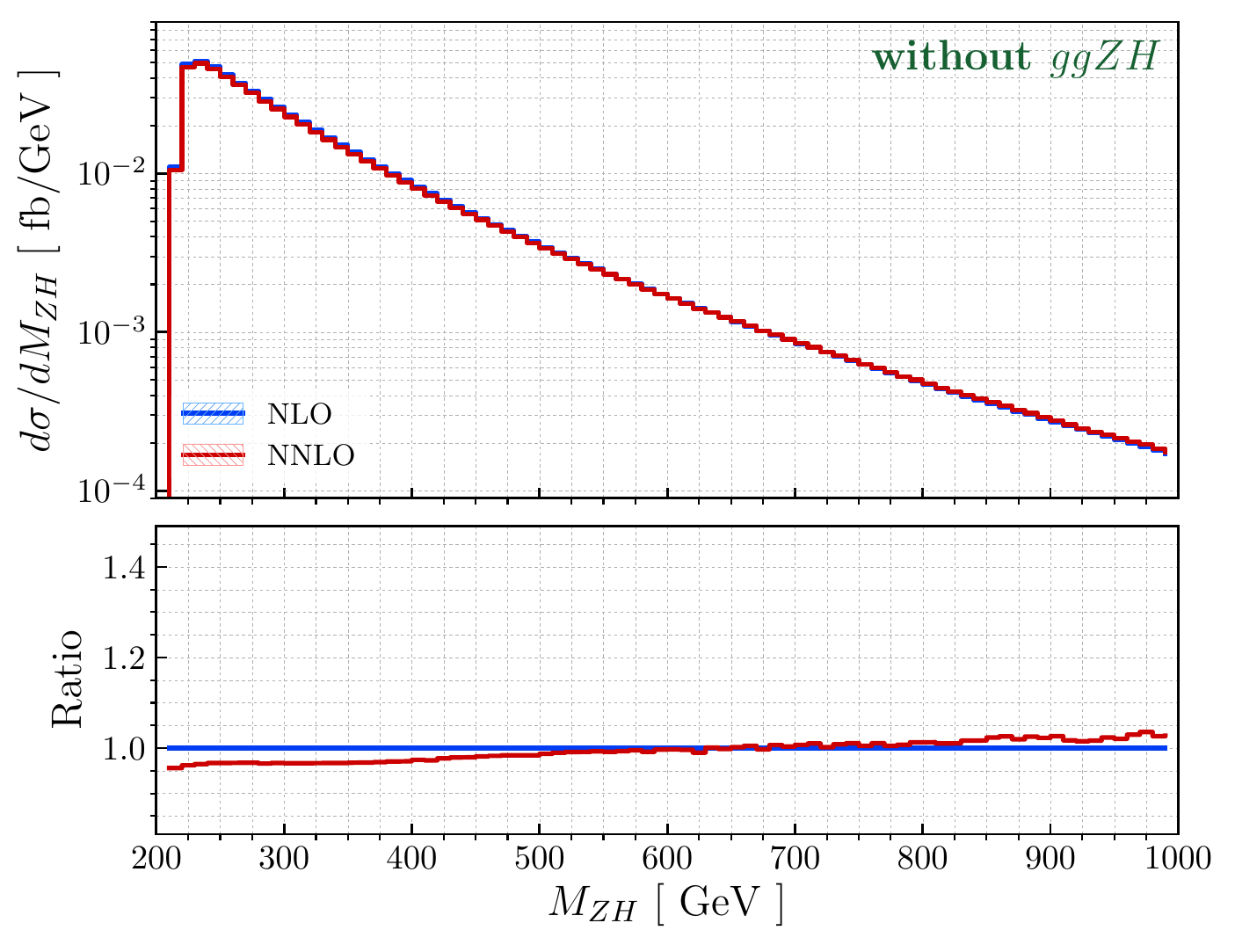}
  \includegraphics[width=0.49\textwidth]{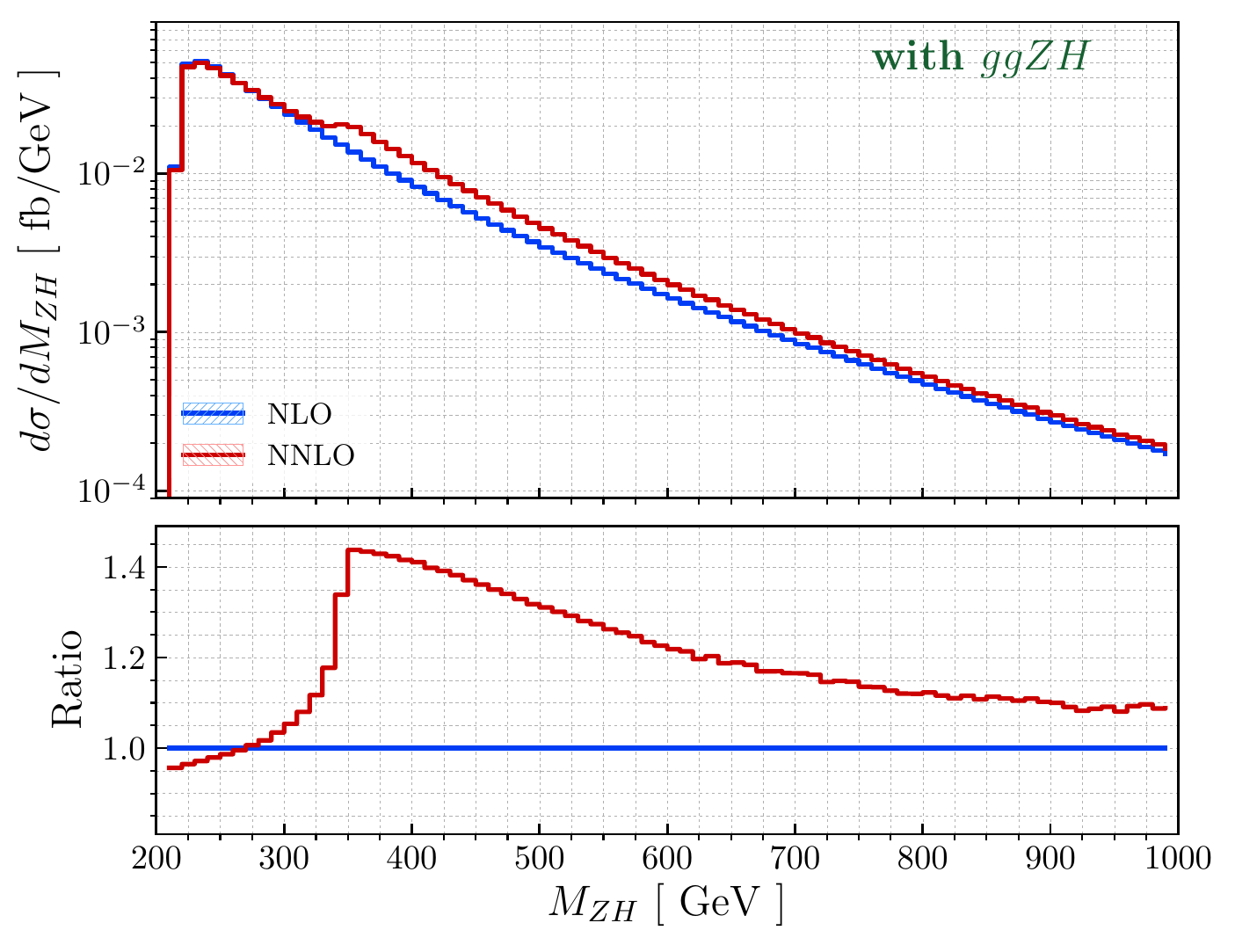}
  \caption{ The invariant mass $M_{ZH}$ distribution at NLO (blue) and
    NNLO (red) at the 13 TeV LHC with the fiducial cuts discussed in the text. We present the NNLO results without (left) and with
    (right) the $gg \to ZH$ contribution. We display results for the central scale $\mu=  \tfrac{1}{2} \sqrt{(p_V+p_H)^2}$.
     The lower panes show the ratio of the NNLO results to the NLO ones. See text for details.}
  \label{fig:MZH}
\end{figure}

Before we turn to the discussion of anomalous couplings, we show a few kinematic  distributions in the Standard Model.
Since we have discussed the $W^+H$ process in some detail earlier~\cite{Behring:2020uzq}, we
focus exclusively on the distributions for the $pp \to ZH$ process.

In Fig.~\ref{fig:MZH} we display the invariant mass of the Higgs boson and  $Z$ boson system. We note that the invariant mass is
reconstructed from the ``true'' Higgs boson momentum $p_H$ and the $Z$-boson momentum $p_Z$, i.e. $M_{ZH}^2 = (p_H + p_Z)^2$;
however,  to be included in the plot, an event is required to pass the kinematic cuts described above.
We observe large changes in  this distribution starting at $M_{ZH} \sim  350~{\rm GeV}$  where the $gg \to ZH$ contribution
becomes significant. However, the NNLO QCD corrections to the quark-initiated  processes change the NLO QCD
distribution only slightly; they are about $-5\%$ at low invariant masses  and  become slightly positive at larger $M_{ZH}$ values.

\begin{figure}\centering
  \includegraphics[width=0.49\textwidth]{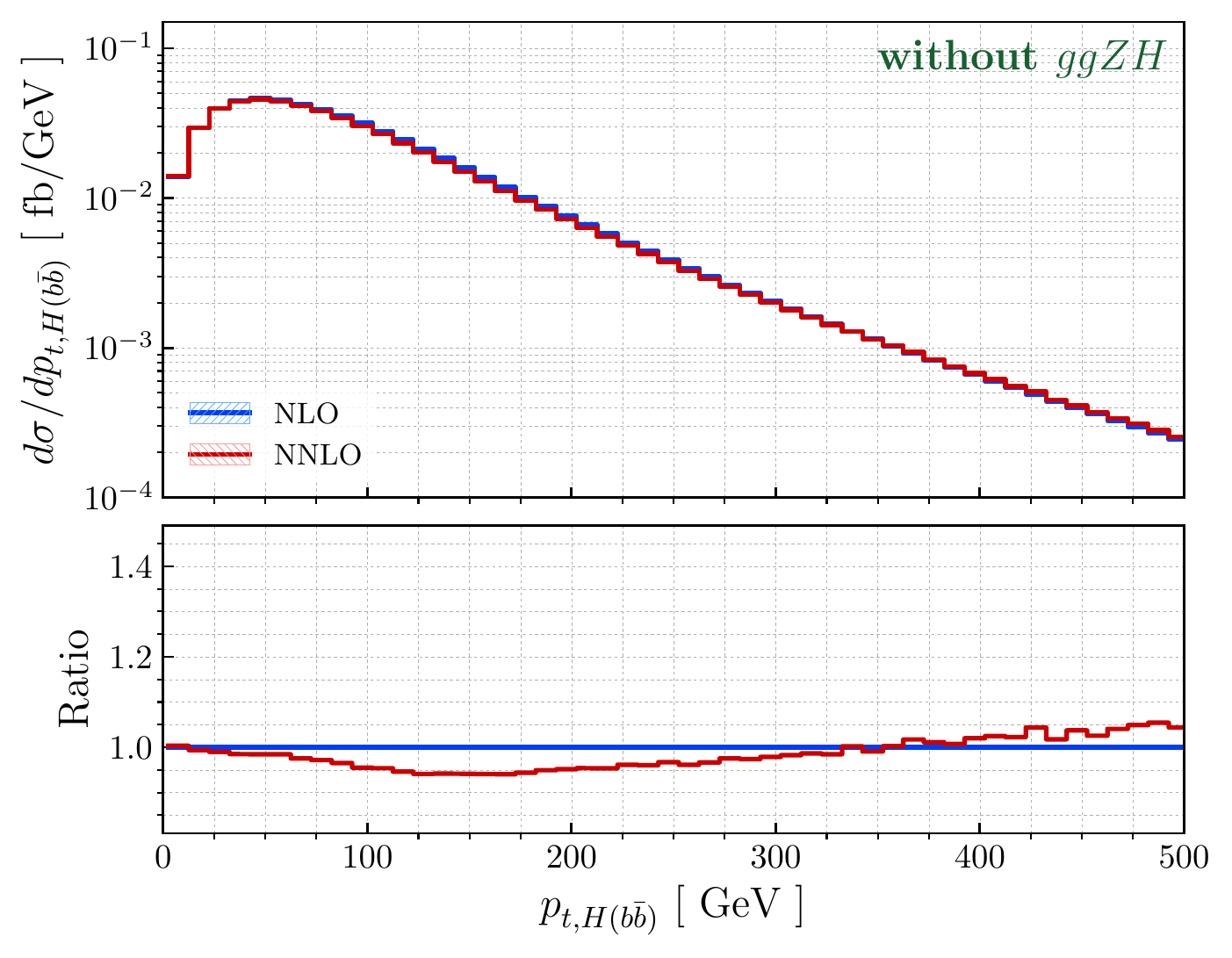}
  \includegraphics[width=0.49\textwidth]{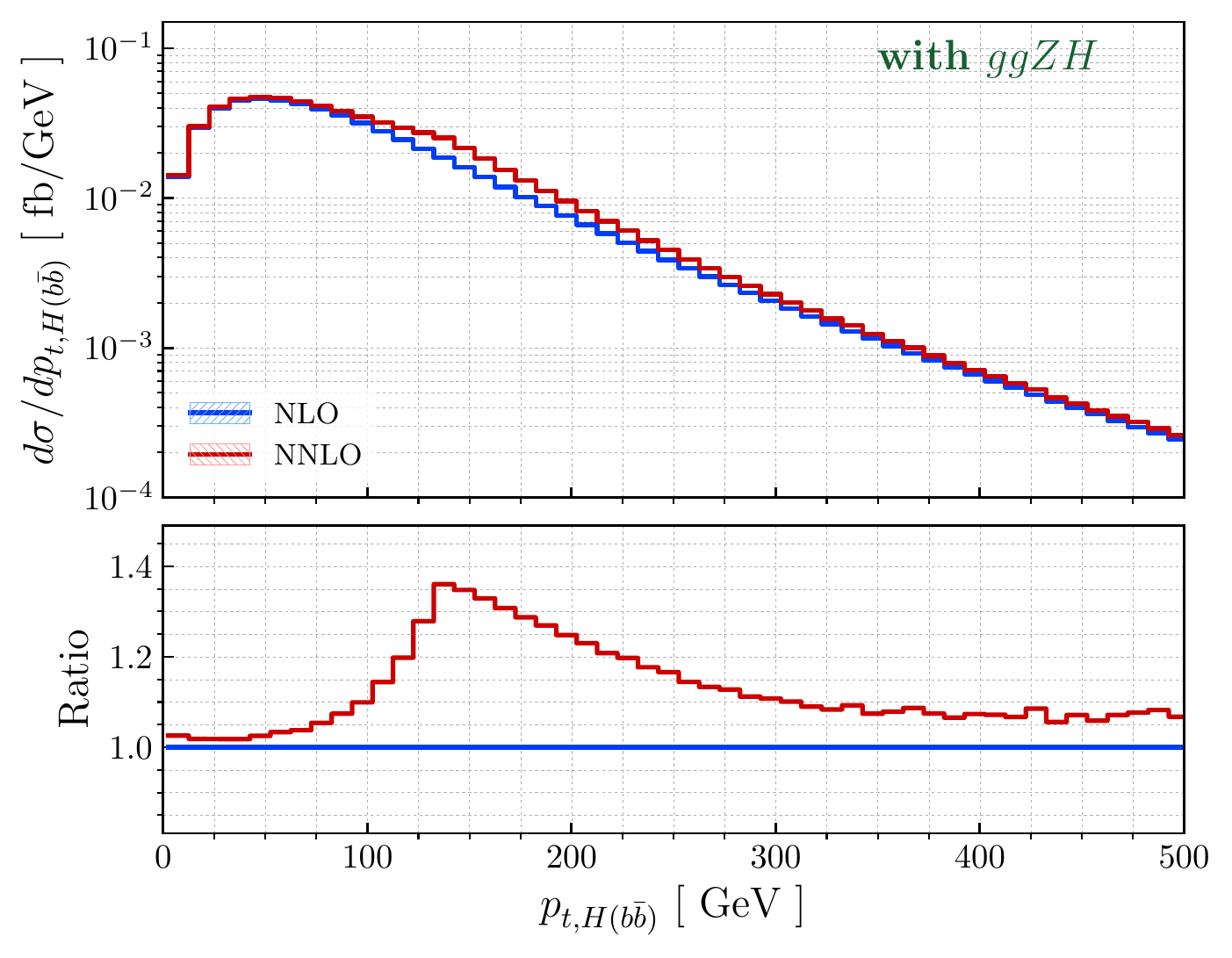}
  \caption{ The transverse momentum distribution of the reconstructed
    Higgs boson at NLO (blue) and NNLO (red) at the 13 TeV LHC with the fiducial cuts discussed in the text. We present the NNLO
    results without (left) and with (right) the $gg \to ZH$ contribution. We display results for the central scale $\mu=  \tfrac{1}{2} \sqrt{(p_V+p_H)^2}$.
     The lower panes show the ratio of the NNLO results to the NLO ones. See text for details.
  }
  \label{fig:pth_rec}
\end{figure}

\begin{figure}\centering
  \includegraphics[width=0.49\textwidth]{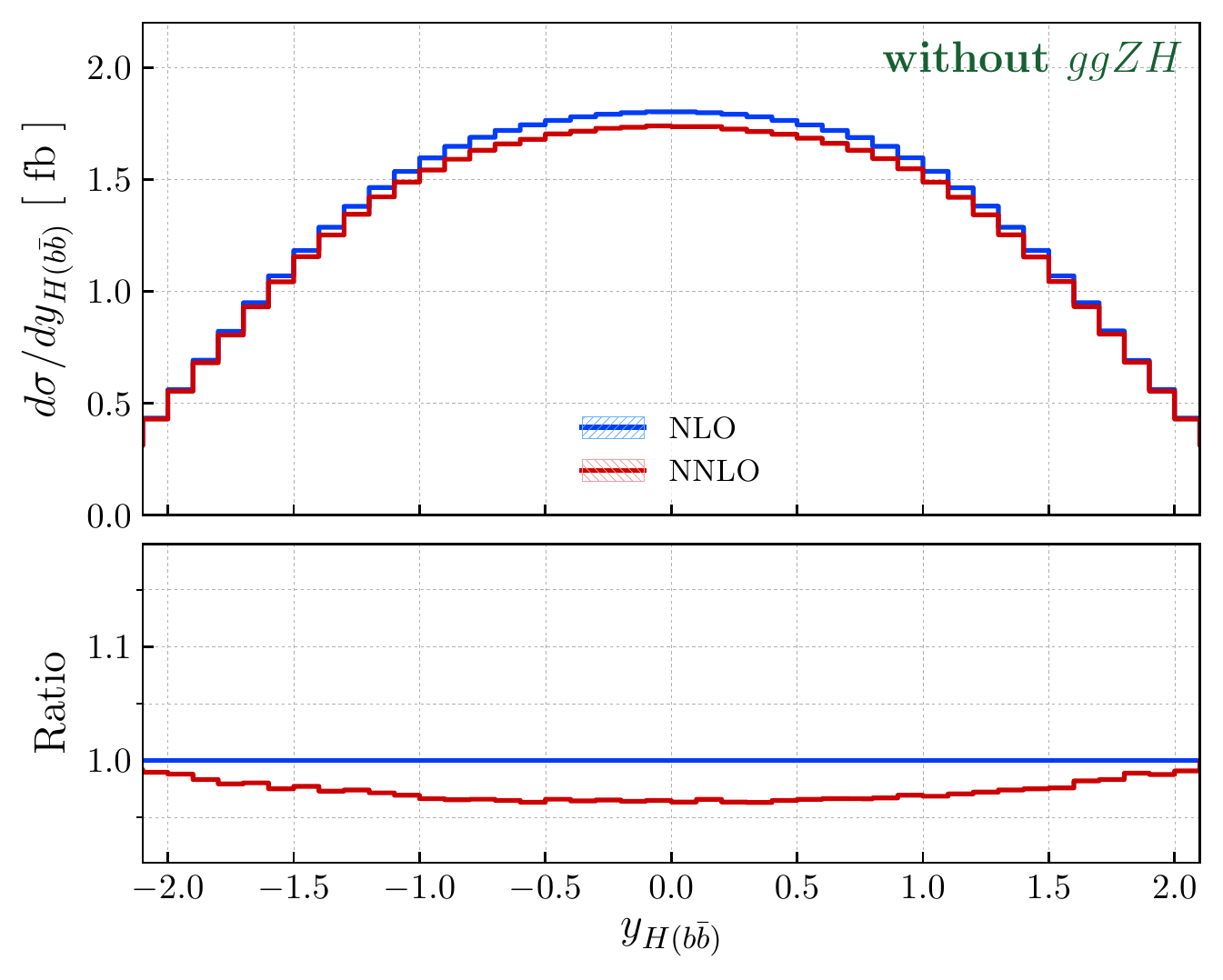}
  \includegraphics[width=0.49\textwidth]{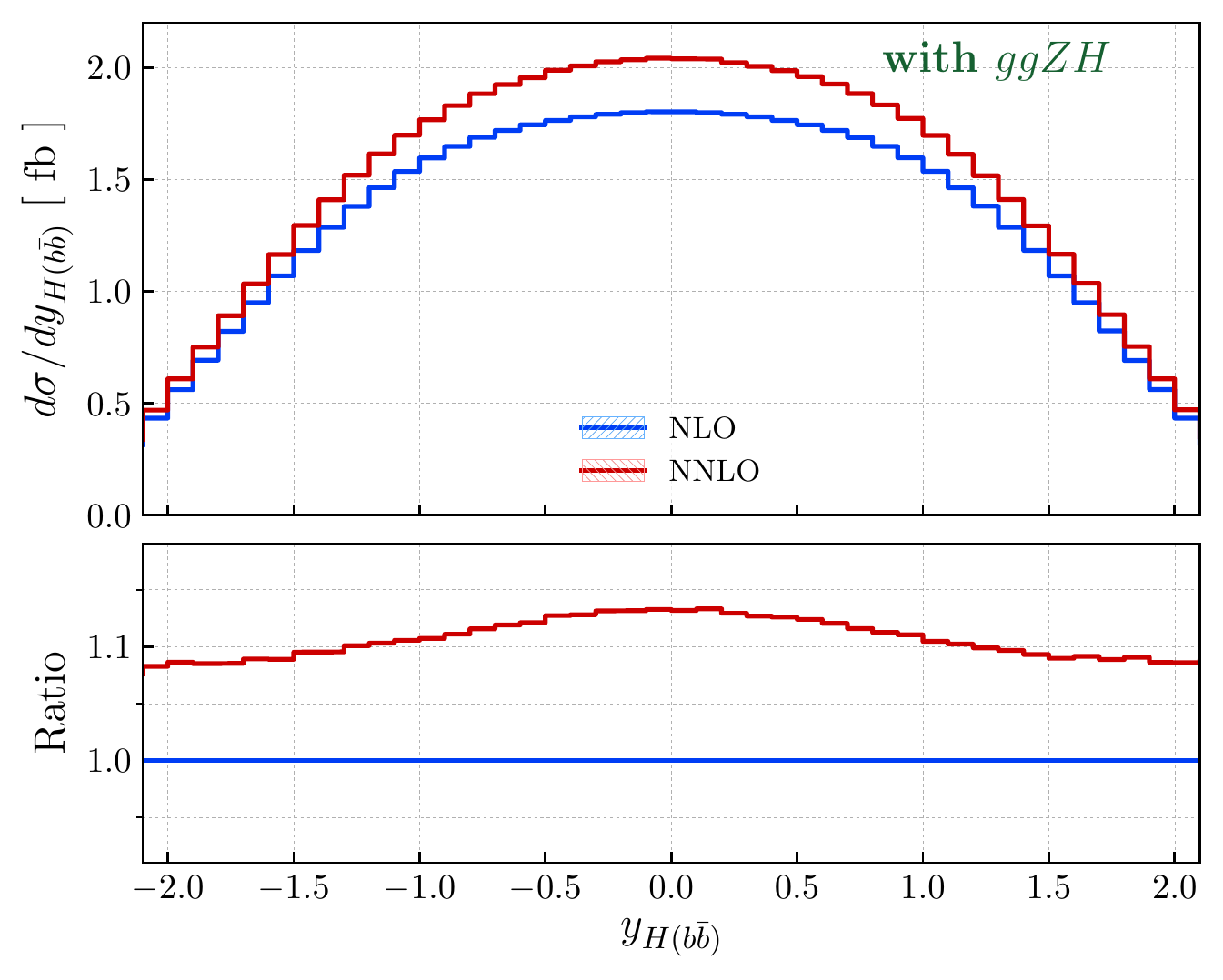}
  \caption{ The rapidity distribution of the reconstructed Higgs boson
    at NLO (blue) and NNLO (red) at the 13 TeV LHC with the fiducial cuts discussed in the text. We present the NNLO results without
    (left) and with (right) the $gg \to ZH$ contribution. We display results for the central scale $\mu=  \tfrac{1}{2} \sqrt{(p_V+p_H)^2}$.
     The lower panes show the ratio of the NNLO results to the NLO ones. See text for details. }
  \label{fig:yh_rec}
\end{figure}

The  transverse momentum and rapidity
distributions of the  Higgs boson are shown in  Figs.~\ref{fig:pth_rec} and~\ref{fig:yh_rec}, respectively.
In these plots, the Higgs momentum is reconstructed from two  $b$ jets as described earlier.
We note that if  more than two $b$ jets appear in the final state, we choose the pair whose invariant mass is closest
to the Higgs boson mass $M_H = 125~{\rm GeV}$.
The corrections to the Higgs transverse momentum distribution show a
pattern that is similar to what is seen in the $ZH$ invariant mass
distribution. Indeed, in the case of quark-initiated $ZH$ production,
the NNLO QCD corrections are negative and decrease NLO distributions
by no more than five percent, whereas if $gg \to ZH$ is included in
the theoretical prediction, there are large modifications of the
$p_{t,H}$ distribution starting at $p_{t,H} \sim 150~{\rm GeV}$.

Kinematic distributions that are integrated over the Higgs boson transverse momentum and $ZH$ invariant masses do not exhibit
local enhancements due to the $gg \to ZH$ process but, rather,  show an  overall increase similar to fiducial cross sections.
This is the case for e.g. the (reconstructed)  Higgs  rapidity distribution shown in Fig.~\ref{fig:yh_rec}; we observe there
that with or without
$gg \to ZH$ contributions, the rapidity distribution is modified by an almost constant $K$-factor.

\begin{figure}\centering
  \includegraphics[width=0.49\textwidth]{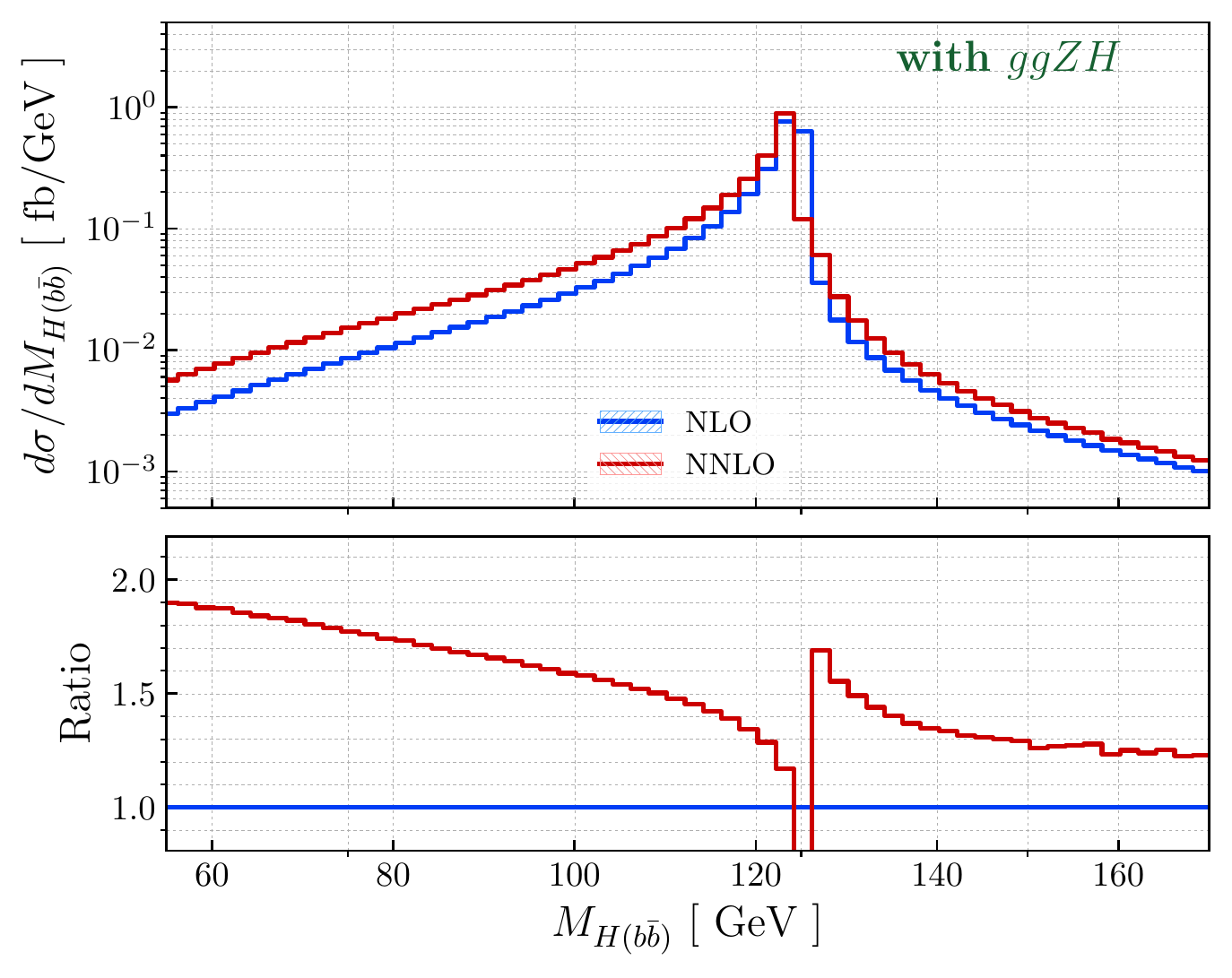}
  \includegraphics[width=0.49\textwidth]{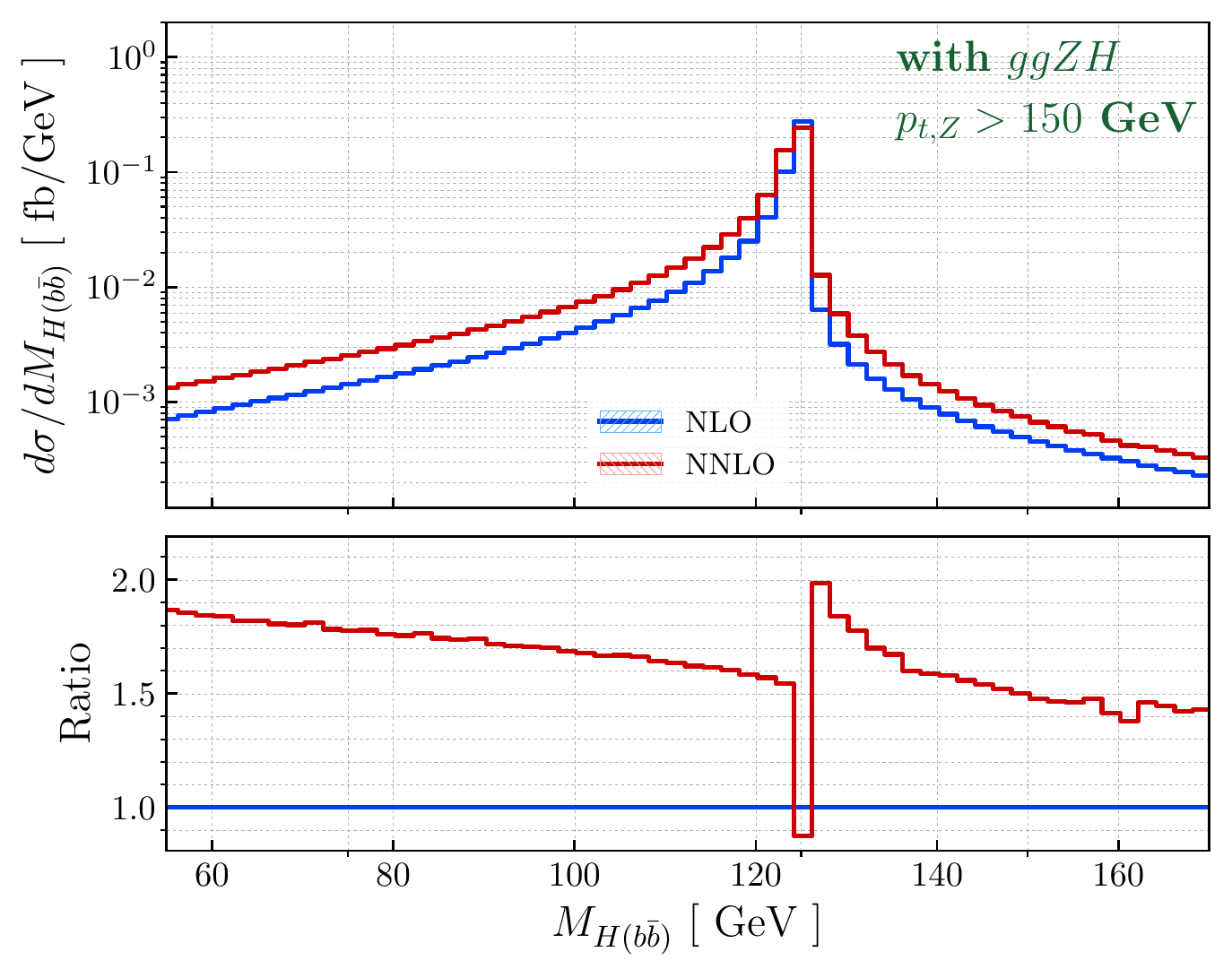}
  \caption{ The invariant mass distribution of the reconstructed Higgs
    boson at NLO (blue) and NNLO (red)  at the 13 TeV LHC. We present the NNLO results
    including the $gg \to ZH$ contribution. The left plot includes the standard
    fiducial cuts described in the text, the right plot includes the 
    additional $\ptVcut$ cut.
    We display results for the central scale $\mu=  \tfrac{1}{2} \sqrt{(p_V+p_H)^2}$.
    The lower panes show the ratio of the NNLO results to the NLO ones.
    See text for details.
  }
  \label{fig:mbb_rec}
\end{figure}

Finally, we give an example of a kinematic distribution that exhibits large NNLO QCD corrections that are not related
to the $gg \to ZH$ process. In Fig.~\ref{fig:mbb_rec} we display the Higgs boson invariant mass distribution where
the Higgs boson is reconstructed from two  $b$ jets whose invariant
mass is the closest to Higgs boson mass.
The $gg \to ZH$ process  contributes only to the $M_{H(b \bar b)} = M_H$ bin
since it has {\it at most} two $b$ jets in the final state and the invariant mass of these $b$ jets is equal to $M_H$.
However, similar to the case of the $W^+H$ final state discussed
in Ref.~\cite{Behring:2020uzq}, we observe very large NNLO QCD effects in the $M_{H(b \bar b)}$ distributions  away from the peak
at the true mass of the Higgs boson due to initial- and final-state radiation.

\section{Associated $VH$ production with anomalous couplings}
\label{sec:anom}

Theoretical predictions for the associated production processes can be
modified by both higher-order QCD effects and by contributions of
physics beyond the Standard Model (BSM).  Under certain circumstances, the
latter can be described by an effective Lagrangian that parameterizes
possible deviations from the Standard Model in terms of operators with
increasing mass dimensions. A convenient description is provided by
the so-called Standard Model Effective Field Theory (SMEFT), see
Ref.~\cite{Brivio:2017vri} for a review.

In  this paper, we will follow Ref.~\cite{Mimasu:2015nqa} and only consider operators that
modify interactions of the Higgs boson to electroweak gauge bosons.  The part of the SMEFT Lagrangian  that is relevant to us
reads
\be
\begin{split}
  \mathcal{L}_{\rm anom}
  ={}&
  -\frac{1}{4\Lambda} g_{  hzz}^{(1)} Z_{\mu\nu} Z^{\mu\nu} h
  -  \frac{1}{2\Lambda} g_{  hww}^{(1)} W^{\mu \nu} W^\dag_{\mu\nu} h
   \\
  &
  -\frac{1}{\Lambda} g_{  hzz}^{(2)} Z_\nu \partial_\mu Z^{\mu\nu} h
  - \frac{1}{\Lambda} \Big[g_{  hww}^{(2)} W^\nu \partial^\mu W^\dag_{\mu\nu} h + {\rm h.c.} \Big]
  \\
  &
  -\frac{1}{4\Lambda} \tilde g_{  hzz} Z_{\mu\nu} \tilde Z^{\mu\nu} h
  -\frac{1}{2\Lambda} \tilde g_{  hww} W^{\mu\nu} \tilde W^\dag_{\mu\nu} h
  \,.
\label{efflag}
\end{split}
  \ee
  The energy scale associated with this Lagrangian is denoted by $\Lambda$; in what follows we will set $\Lambda$ to $1~{\rm TeV}$ for
  definiteness.
  Parametrically, modifications of the Standard Model predictions due to operators in Eq.~\eqref{efflag}
  are controlled by the quantities $g^{(i)}_{hVV}\; v/\Lambda$ where $v= 246~{\rm GeV}$ is the Higgs field
  vacuum expectation value. In what follows, we will consider values of the couplings that lead to relatively small deviations from Standard Model
  predictions and discuss to what extent  better quality  theoretical predictions for the associated production processes $pp \to VH$
  may help with detecting and analyzing such scenarios.

It is straightforward
to incorporate    effects of the anomalous couplings into theoretical predictions for cross sections and kinematic distributions.
To this end, we note that the above Lagrangian   leads to the following  $HV(q_1)V(q_2)$ interaction vertex 
\be
-g^{\mu \nu} c_1 + c_2 \left ( q_1^\mu q_2^\nu + q_1^\nu q_2^\mu \right )
+ c_3 \epsilon^{\mu \nu \alpha \beta} q_{1,\alpha} q_{2,\beta} + c_4 \left ( q_1^\mu q_1^\nu + q_2^\mu q_2^\nu \right ).
\label{eqhvv}
\ee
In Eq.~\eqref{eqhvv} the coefficient  $c_1$ is a first-degree  polynomial in $q_1^2$,   $q_2^2$ and $q_1 q_2$,   whereas coefficients
$c_{2,3,4}$ are independent of the external momenta. Also,  the coefficients $c_{1,..,4}$
are functions of the various couplings $g_{hVV}$ in the Lagrangian Eq.~(\ref{efflag}); the exact relations between
the coefficients $c_{1..4}$ and the various $g_{hVV}$ couplings  can be found in Fig. 1 of Ref.~\cite{Mimasu:2015nqa} and we do not reproduce  them here.

As we already mentioned in Section~\ref{sec:sm},  analytic formulas for the integrated  subtraction terms required for the NNLO QCD
description of color-singlet production~\cite{Caola:2019nzf}  are generic.
For this reason all that we need to do in order to incorporate  effects of the anomalous
couplings into a NNLO QCD description of the $pp \to VH$ process is to provide scattering amplitudes for hard processes
$q\bar q' \to H l_1 \bar l_2$,  $q\bar q' \to H l_1 \bar l_2+g$, $q\bar q' \to H l_1 \bar l_2+gg$ {\rm etc.} that include the anomalous couplings.
Once these amplitudes are available -- and it is quite straightforward to calculate them -- they  can be immediately included
in a numerical code for computing cross sections and kinematic distributions for the associated production processes through
NNLO in perturbative QCD.

It is to be expected that generic choices of anomalous couplings would
lead to significant changes in cross sections and kinematic
distributions.  For such cases an extraction of the values of the
anomalous couplings from data, rather than the detection of anomalies,
would benefit from precise predictions for observables that include
the anomalous couplings.  On the other hand, there are also cases
where, even with anomalous couplings, changes in cross sections
are marginal. In this situation, studies of kinematic distributions
and precise theoretical predictions may be needed to both detect the
presence of anomalies and distinguish between different scenarios.  In
what follows we present a few examples.

We have seen in the previous section that existing predictions for the $gg \to ZH$ partonic process are insufficiently precise, leaving
up to  six percent uncertainty in predictions for fiducial cross sections.  For this reason, it is desirable to reduce the impact of this contribution.
Since the relevance of the $gg \to ZH$ channel grows at high invariant masses of the $ZH$ system,  putting an upper   kinematic cut on $M_{ZH}$ is useful to
increase the quality of theoretical predictions without reducing fiducial cross sections. At the same time, an upper kinematic cut on $M_{ZH}$
has the additional benefit of removing contributions of high-energy tails of distributions where an EFT expansion may become unreliable.   We note that
since it is not possible to impose such a cut on the $W^+H$ system, we also  restrict the transverse momentum of the vector boson following the 
experimental analysis~\cite{Aad:2020jym}.

Hence,   we will study   fiducial cross sections and
kinematic distributions of the associated production processes $pp \to VH$ including the anomalous couplings
by imposing the kinematic cuts of Eq.~(\ref{eq:cuts}) as well as the following constraints:
\begin{align}
  \label{eq:zh-fid-anom}
  ZH:\;\;\;\;\;  &  75~\gev < p_{t,Z} < 250~\gev,\;\;\;\;M_{e^+ e^- b \bar b} < 320~\gev, \\
  \label{eq:wh-fid-anom}
  W^+H:\;\;\;\;\;  & 150~\gev < p_{t,W} < 250~\gev.
\end{align}
The notation $M_{e^+ e^- b \bar b}$
emphasizes the fact that the invariant mass of the $ZH$ system  is
calculated using the four-momenta of the two  charged leptons and the two
$b$ jets used for the Higgs boson  reconstruction.

\subsection{$ZH$ process}

We begin with the discussion of the $pp \to ZH$ process and consider the following scenarios:
\begin{align*}
  \text{Setup 1: } \qquad&
  g_{hzz}^{(1)}   ={} + 2.80\,, \quad
  g_{hzz}^{(2)}   ={} - 0.60\,, \quad
  \tilde{g}_{hzz} ={} + 0.00\,, \\
  \text{Setup 2: } \qquad&
  g_{hzz}^{(1)}   ={} + 1.05\,, \quad
  g_{hzz}^{(2)}   ={} + 0.00\,, \quad
  \tilde{g}_{hzz} ={} - 2.90\,, \\
  \text{Setup 3: } \qquad&
  g_{hzz}^{(1)}   ={} + 0.00\,, \quad
  g_{hzz}^{(2)}   ={} - 1.00\,, \quad
  \tilde{g}_{hzz} ={} - 3.30\,, \\
  \text{Setup 4: } \qquad&
  g_{hzz}^{(1)}   ={} + 1.00\,, \quad
  g_{hzz}^{(2)}   ={} - 1.00\,, \quad
  \tilde{g}_{hzz} ={} + 2.00\,.
\end{align*}

\begin{table}[t]
  \centering
  \small
  \begin{tabular}{lccccc}

    \hline
    $\sigma^{ZH}_{\rm fid}$ $[\fb]$
    & SM
    & Setup 1
    & Setup 2
    & Setup 3
    & Setup 4
    \\

    \hline
    $\LO$
    &  $ 0.894^{ +0.032 }_{ -0.041 }  $  
    &  $ 0.782^{ +0.028 }_{ -0.034 }  $  
    &  $ 0.854^{ +0.031 }_{ -0.038 }  $  
    &  $ 0.786^{ +0.027 }_{ -0.034 }  $  
    &  $ 0.780^{ +0.027 }_{ -0.034 }  $  
    \\

    \hline
    $\NLO$
    &  $ 1.289^{ +0.025 }_{ -0.017 }  $  
    &  $ 1.266^{ +0.012 }_{ -0.007 }  $  
    &  $ 1.273^{ +0.018 }_{ -0.010 }  $  
    &  $ 1.276^{ +0.009 }_{ -0.004 }  $  
    &  $ 1.269^{ +0.008 }_{ -0.003 }  $  
    \\

    \hline
    $\NNLO$
    &  $ 1.356^{ +0.009 }_{ -0.011 }  $  
    &  $ 1.423^{ +0.003 }_{ -0.006 }  $  
    &  $ 1.379^{ +0.014 }_{ -0.004 }  $  
    &  $ 1.454^{ +0.003 }_{ -0.006 }  $  
    &  $ 1.445^{ +0.004 }_{ -0.003 }  $  
    \\

    \hline
    ${\NNLO}$($+ggZH$)
    &  $ 1.419^{ +0.024 }_{ -0.023 }  $  
    &  $ 1.476^{ +0.015 }_{ -0.015 }  $  
    &  $ 1.443^{ +0.028 }_{ -0.015 }  $  
    &  $ 1.499^{ +0.014 }_{ -0.015 }  $  
    &  $ 1.490^{ +0.014 }_{ -0.011 }  $  
    \\

    \hline

  \end{tabular}
  \caption{Fiducial cross sections  for $pp \to ZH \to (e^-e^+)(\bb)$ at the $13~{\rm TeV}$ LHC at various orders of
    QCD perturbation theory calculated with massive $b$ quarks. We show the results for various scenarios including anomalous
    couplings.
    We set the factorization and renormalization scales equal to each other,  $\mu_r = \mu_f = \mu$.
    We use $\mu=  \tfrac{1}{2} \sqrt{(p_V+p_H)^2}$ for the central value and the uncertainties are calculated by varying the scale $\mu$ by a factor of two in both directions. 
    See text for details.}
  \label{tab:anom-xsec-zh}
\end{table}

Although these choices look quite random, the corresponding scenarios were chosen to provide almost identical cross sections both at leading and, especially,
at next-to-leading order in QCD,
subject to the kinematic constraints shown in Eq.~\eqref{eq:cuts}
and Eq.~\eqref{eq:zh-fid-anom}.
This can be clearly seen from the results for fiducial cross sections  summarized  in Table~\ref{tab:anom-xsec-zh}.
We  observe that NNLO QCD corrections in these cases are not insignificant; they lead to  important shifts compared to next-to-leading order
predictions.  Also, the very strong degeneracy of the four scenarios at NLO is lifted at NNLO.
However, the differences between predictions
for the different scenarios remain within a few percent of each other, making NNLO QCD precision for these cases essential.
It also follows from  Table~\ref{tab:anom-xsec-zh} that for such situations
it is important to include higher-order QCD corrections to the description of processes
with anomalous couplings since simply  re-weighting the  leading order predictions with Standard Model
$K$-factors may be insufficient.
At any rate, having  high-precision predictions  for fiducial cross sections
of processes with anomalous couplings
may help with analyzing cases where differences between various  scenarios are marginal.

\begin{figure}[t]\centering
  \includegraphics[width=0.49\textwidth]{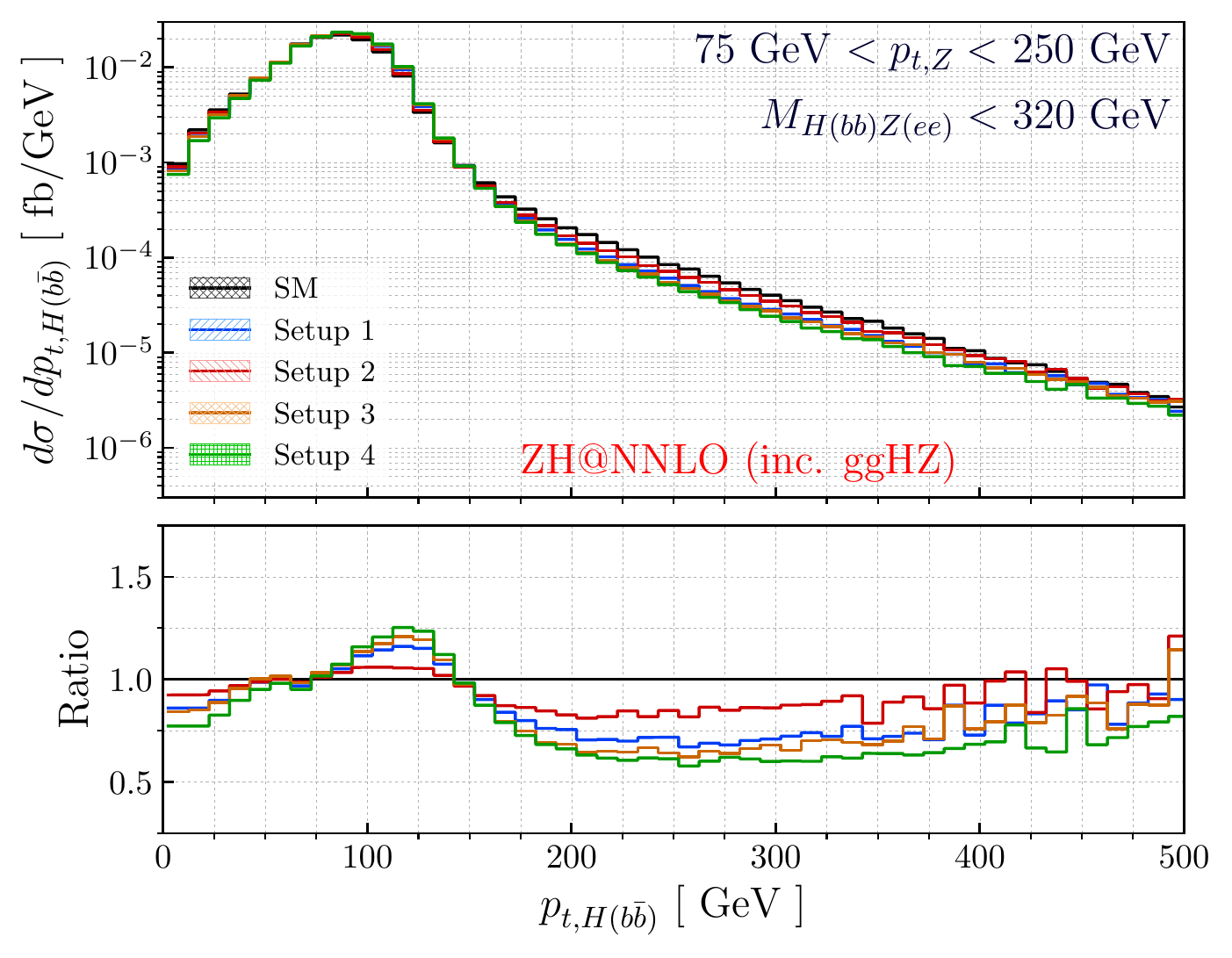}
  \includegraphics[width=0.49\textwidth]{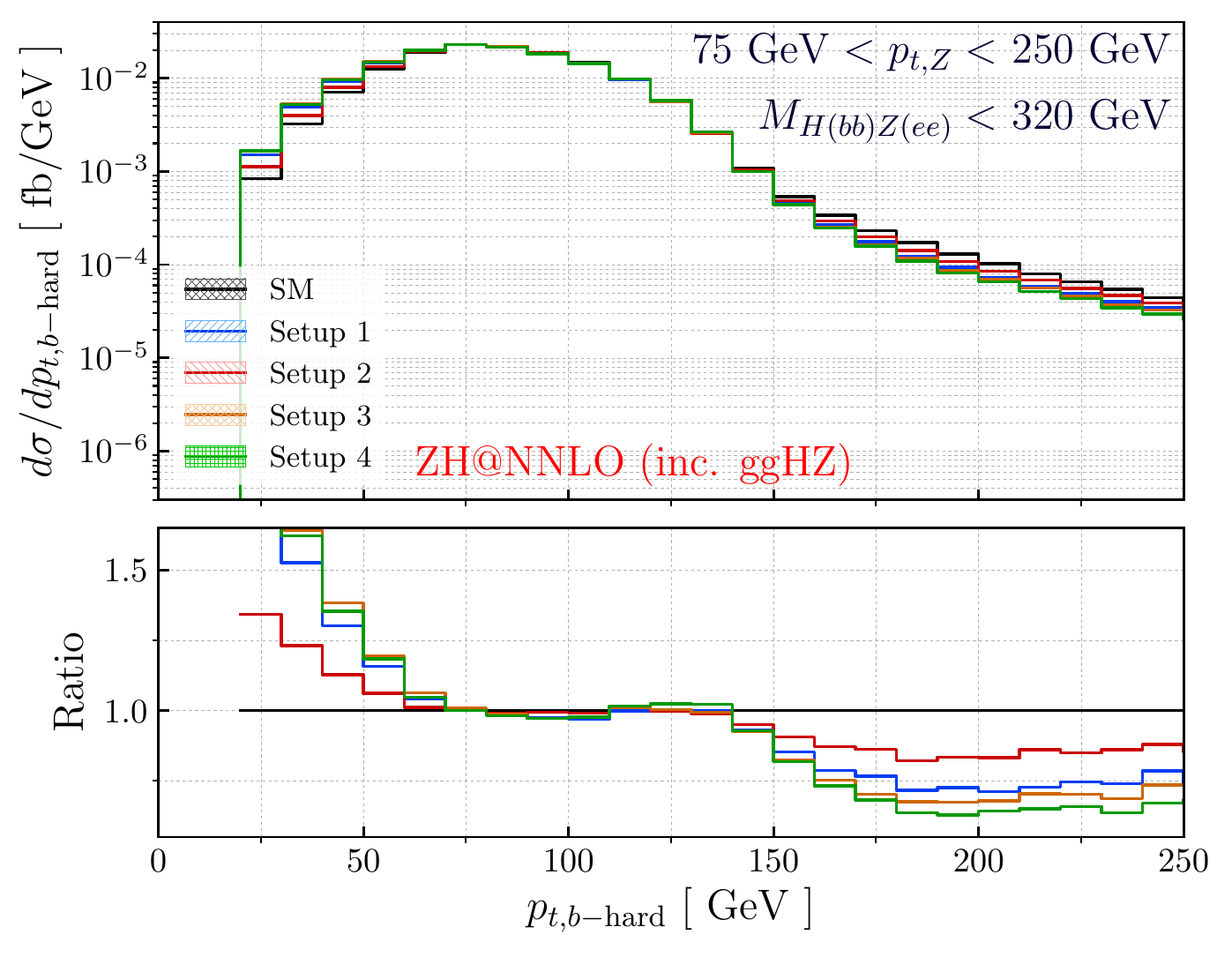}
    \includegraphics[width=0.49\textwidth]{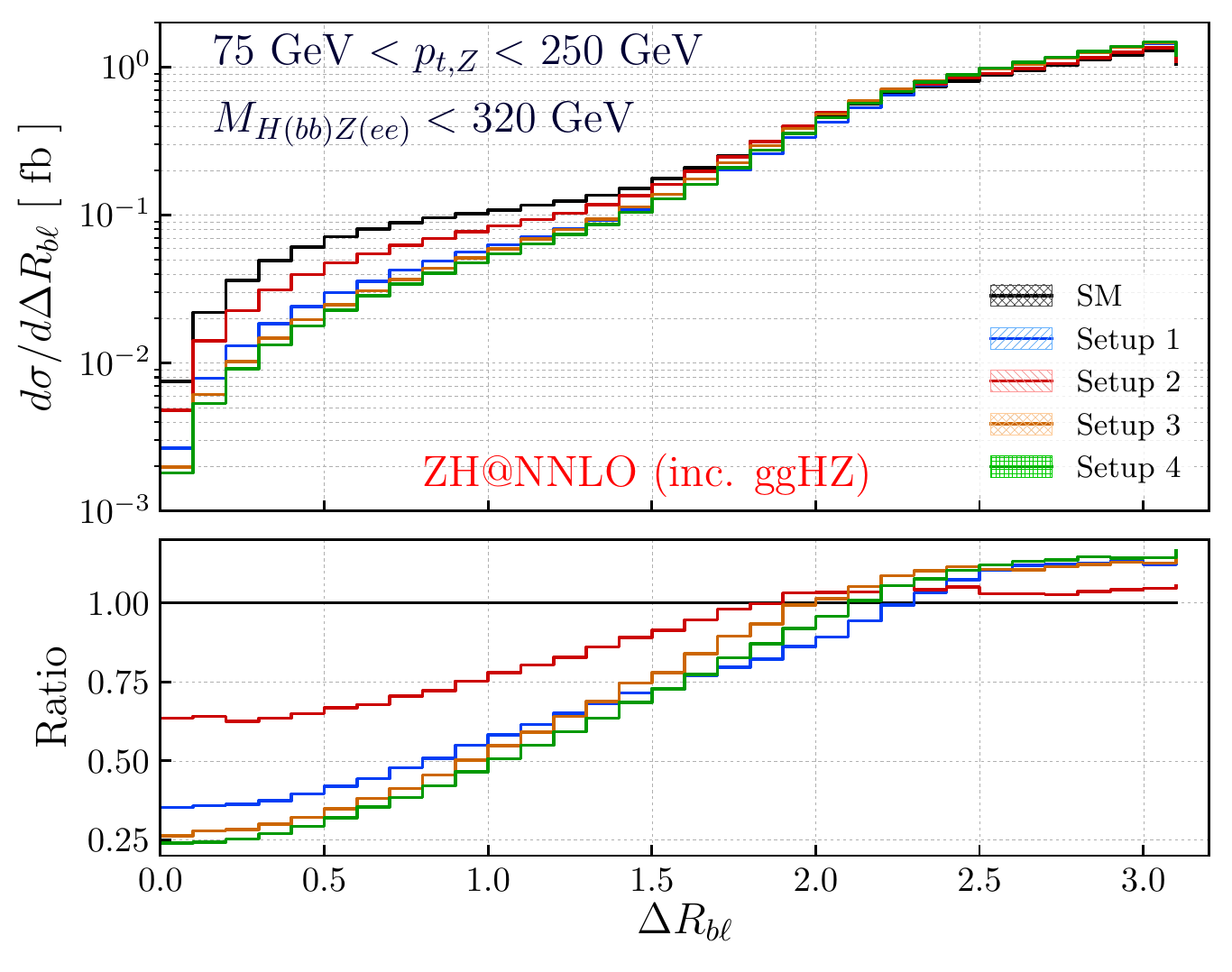}
    \includegraphics[width=0.49\textwidth]{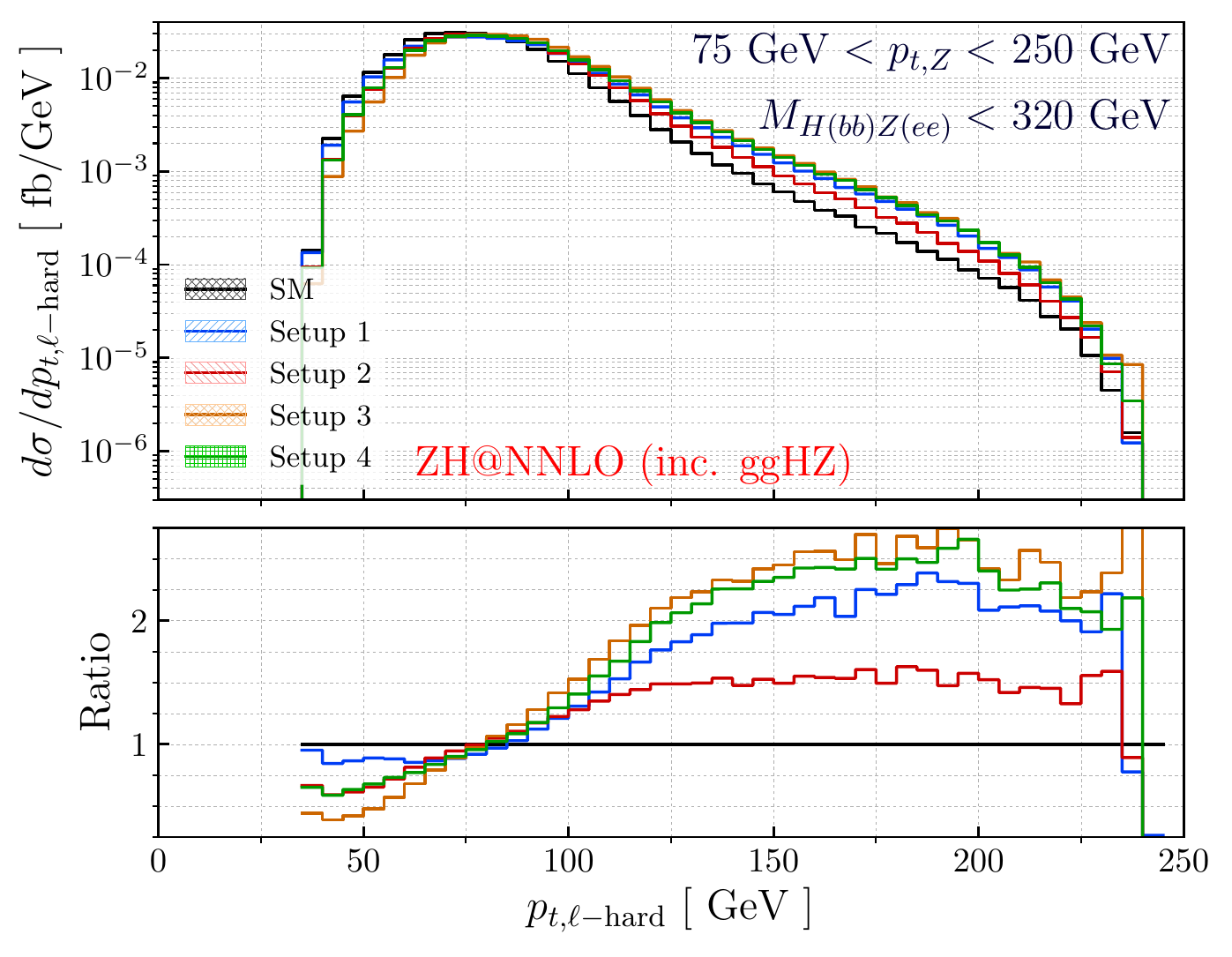}
  \caption{ Kinematic distributions in the process $pp \to
    Z(e^+e^-)H(b \bar b)$ at the 13 TeV LHC for various SMEFT
    scenarios.  In the lower panes, ratios of SMEFT to SM
    distributions are shown.  We set the factorization and the
    renormalization scales in the production process to half the
    invariant mass of the $ZH$ system.  See text for 
    details. }
  \label{fig:zh_pth_rec_anom}
\end{figure}

%
Another way to lift the degeneracies of different scenarios is to explore kinematic distributions. In many cases kinematic distributions
offer more opportunities to detect the anomalous couplings since their effects can be quite profound even if they are small in fiducial
cross sections.  However, for the four scenarios that we considered, the situation is slightly more subtle, as we will illustrate now.

In Fig.~\ref{fig:zh_pth_rec_anom} we show distributions of  the Higgs boson  transverse momentum,
the transverse momentum of the hardest $b$ jet, 
the  angular separation between the hardest $b$ jet and the hardest lepton
$\Delta R_{b\ell}$, and the transverse momentum distribution of the hardest lepton.
The quantity $\Delta R_{b\ell}$ is defined as
\begin{align}
  \Delta R_{b\ell}
  ={}&
  \sqrt{
    (y_{b} - y_\ell)^2
    + (\varphi_{b} - \varphi_\ell)^2
  }\,,
\end{align}
where $y_{b}$($y_{\ell}$) and $\varphi_{b}$($\varphi_{\ell}$) are the rapidity and the azimuthal angle  of the hardest $b$ jet (the hardest lepton),  respectively.

It follows from Fig.~\ref{fig:zh_pth_rec_anom} that there are kinematic  regions where the differences between the four scenarios are more pronounced
than in fiducial  cross sections.
For example, if we look at the $p_{t,H(b \bar b)}$ distribution, there are noticeable differences at {\it low}  transverse momenta,
whereas in the peak region all four distributions are similar. The same applies to the other three distributions.
 For example, the $\Delta R_{bl}$ distributions peak at $\Delta R_{bl} \sim 3$; in that region  the four scenarios provide very
similar results. The differences become noticeable at $\Delta R_{bl} \sim 1$ but the number of events for such values
of $\Delta R_{bl}$ is reduced by an order of magnitude.  Given the fact that we deal here with ${\cal O}(1~{\rm fb})$ cross sections, losing an order of magnitude in
the number of events is not optimal.
However, the availability  of highly accurate NNLO QCD predictions in peak regions of kinematic
distributions and identifiable  differences between various scenarios in distribution tails should allow one to optimize
analysis strategies and benefit from measurements across accessible kinematic regions.

\subsection{$W^+H$ process}

We repeat the analysis of the previous subsection for $W^+H$ production.
We focus exclusively on the fiducial region defined in Eq.~\eqref{eq:cuts}
with additional restrictions on  the $W$-boson transverse momentum, shown in  Eq.~\eqref{eq:wh-fid-anom}.

We consider four different scenarios of the anomalous couplings and we choose them in a way that makes the differences between fiducial cross sections
marginal. The four scenarios are:
\begin{align*}
  \text{Setup 1: } \qquad&
  g_{hww}^{(1)}   ={} - 1.20\,, \quad
  g_{hww}^{(2)}   ={} - 0.25\,, \quad
  \tilde{g}_{hww} ={} + 0.00\,, \\
  \text{Setup 2: } \qquad&
  g_{hww}^{(1)}   ={} + 1.00\,, \quad
  g_{hww}^{(2)}   ={} + 0.00\,, \quad
  \tilde{g}_{hww} ={} + 0.80\,, \\
  \text{Setup 3: } \qquad&
  g_{hww}^{(1)}   ={} + 0.00\,, \quad
  g_{hww}^{(2)}   ={} - 0.10\,, \quad
  \tilde{g}_{hww} ={} - 1.10\,, \\
  \text{Setup 4: } \qquad&
  g_{hww}^{(1)}   ={} + 0.70\,, \quad
  g_{hww}^{(2)}   ={} - 0.05\,, \quad
  \tilde{g}_{hww} ={} - 1.05\,.
\end{align*}

\begin{table}[t]
  \centering
  \small
  \begin{tabular}{lccccc}

    \hline
    $\sigma^{W^+H}_{\rm fid}$ $[\fb]$
    & SM
    & Setup 1
    & Setup 2
    & Setup 3
    & Setup 4
    \\

    \hline
    $\LO$
    &  $ 2.813^{ +0.023 }_{ -0.039 }  $  
    &  $ 2.657^{ +0.012 }_{ -0.024 }  $  
    &  $ 2.999^{ +0.007 }_{ -0.021 }  $  
    &  $ 2.898^{ +0.012 }_{ -0.026 }  $  
    &  $ 2.958^{ +0.007 }_{ -0.021 }  $  
    \\

    \hline
    $\NLO$
    &  $ 3.434^{ +0.089 }_{ -0.064 }  $  
    &  $ 3.419^{ +0.110 }_{ -0.080 }  $  
    &  $ 3.466^{ +0.070 }_{ -0.048 }  $  
    &  $ 3.501^{ +0.088 }_{ -0.063 }  $  
    &  $ 3.458^{ +0.074 }_{ -0.052 }  $  
    \\

    \hline
    $\NNLO$
    &  $ 3.409^{ +0.024 }_{ -0.025 }  $  
    &  $ 3.436^{ +0.028 }_{ -0.034 }  $  
    &  $ 3.387^{ +0.004 }_{ -0.015 }  $  
    &  $ 3.463^{ +0.015 }_{ -0.031 }  $  
    &  $ 3.390^{ +0.003 }_{ -0.018 }  $  
    \\

    \hline

  \end{tabular}
  \caption{ Fiducial cross sections  for $pp \to W^+H \to (\nu_e e^+)(\bb)$ at the $13~{\rm TeV}$ LHC at various orders of
    QCD perturbation theory calculated with massive $b$ quarks. We show the results for various scenarios including anomalous
    couplings.
    We set the factorization and renormalization scales equal to each other,  $\mu_r = \mu_f = \mu$.
    We use $\mu=  \tfrac{1}{2} \sqrt{(p_V+p_H)^2}$ for the central value and the uncertainties are calculated by varying the scale $\mu$ by a factor of two in both directions. 
    See main text for details.}
  \label{tab:anom-xsec-wh}
\end{table}

The fiducial cross sections at various orders of perturbation theory
are reported in Table~\ref{tab:anom-xsec-wh}. We observe that the NLO QCD predictions for cross sections
for the four scenarios agree to within a few percent. At variance with $ZH$ case, however, adding NNLO QCD corrections
does not change the situation in a significant way except that the uncertainty on the theoretical predictions is
reduced compared to the NLO QCD case.
However, we again observe that the NNLO QCD corrections are not constant across the four scenarios.

\begin{figure}\centering
  \includegraphics[width=0.49\textwidth]{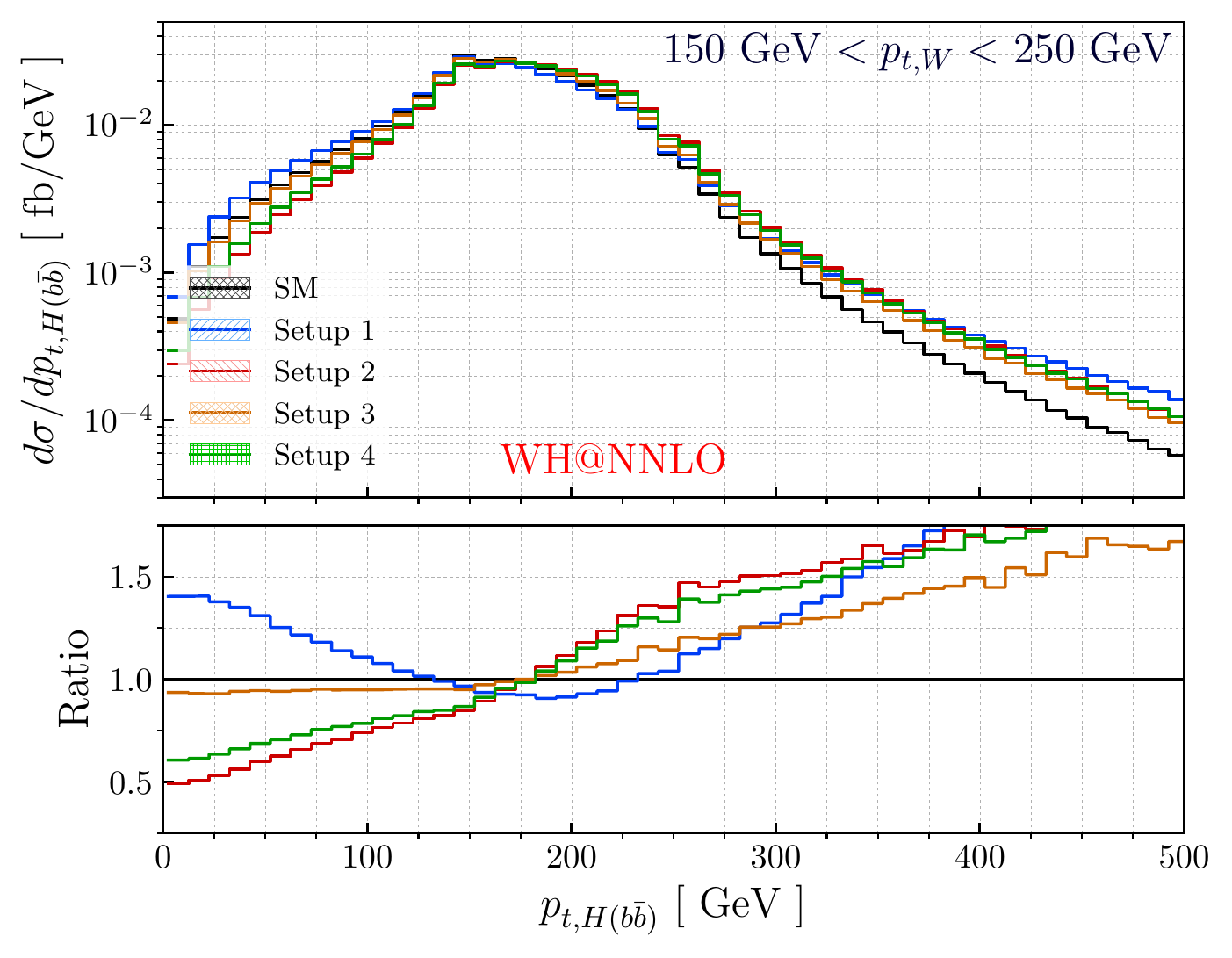}
  \includegraphics[width=0.49\textwidth]{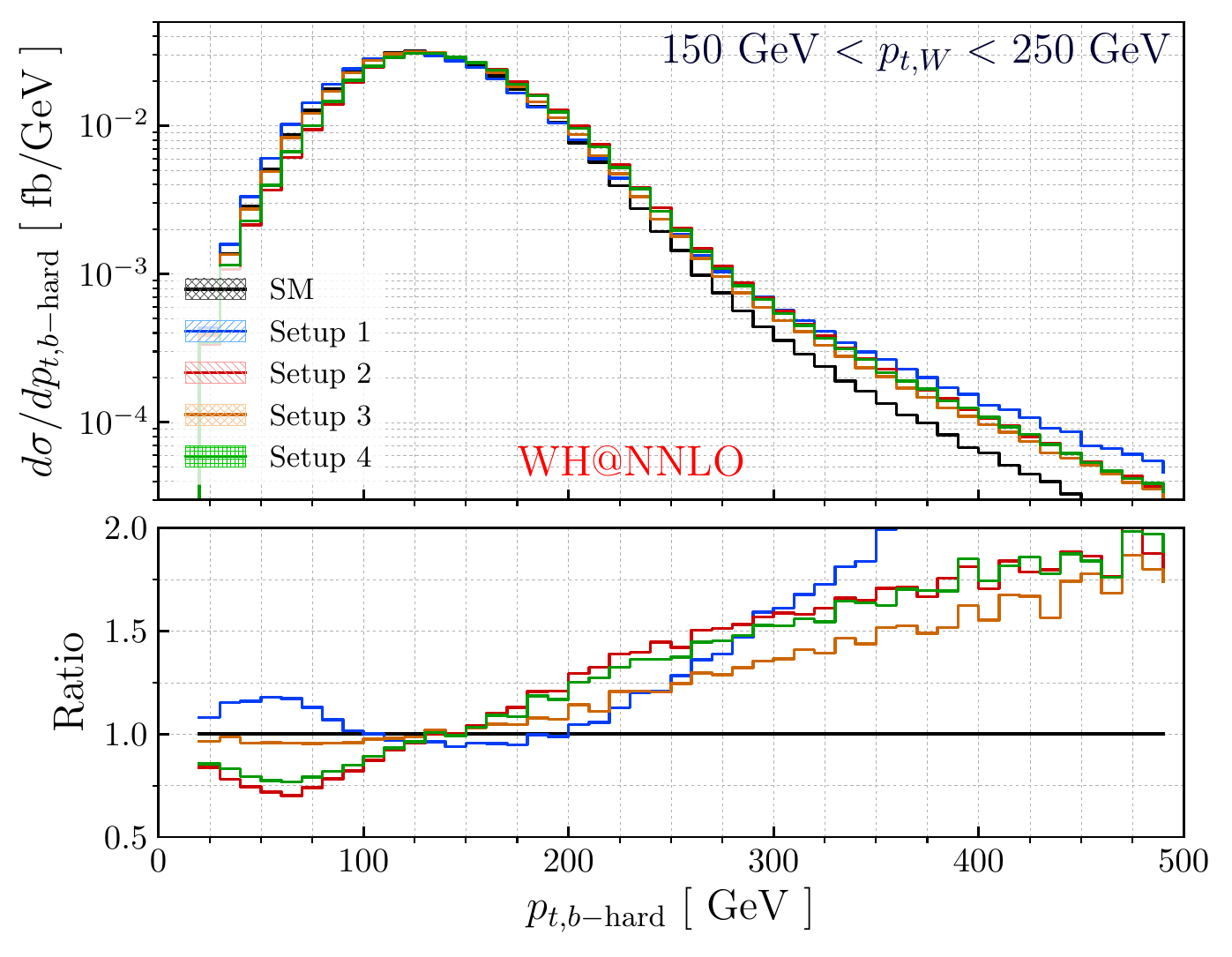}
  \includegraphics[width=0.49\textwidth]{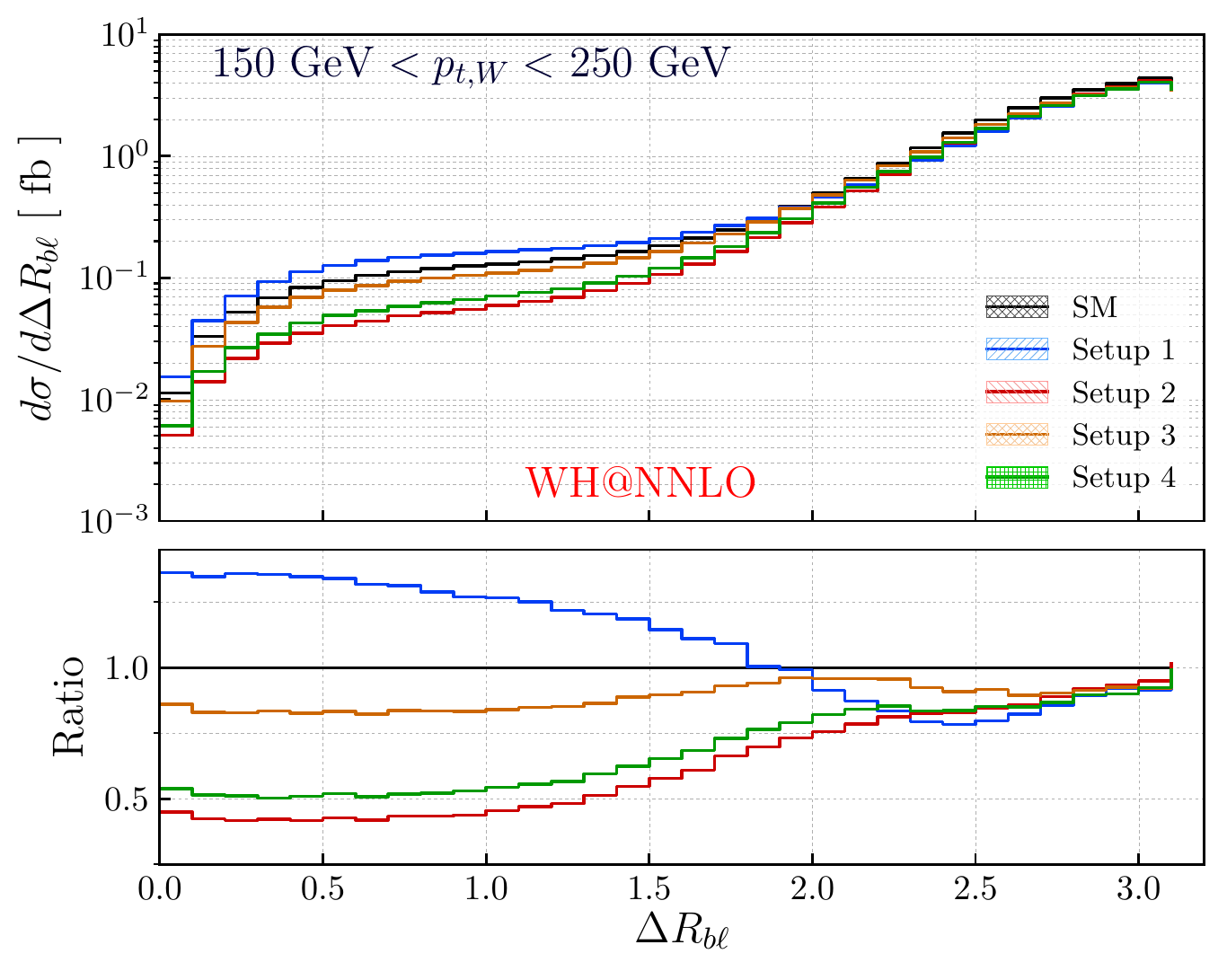}
    \includegraphics[width=0.49\textwidth]{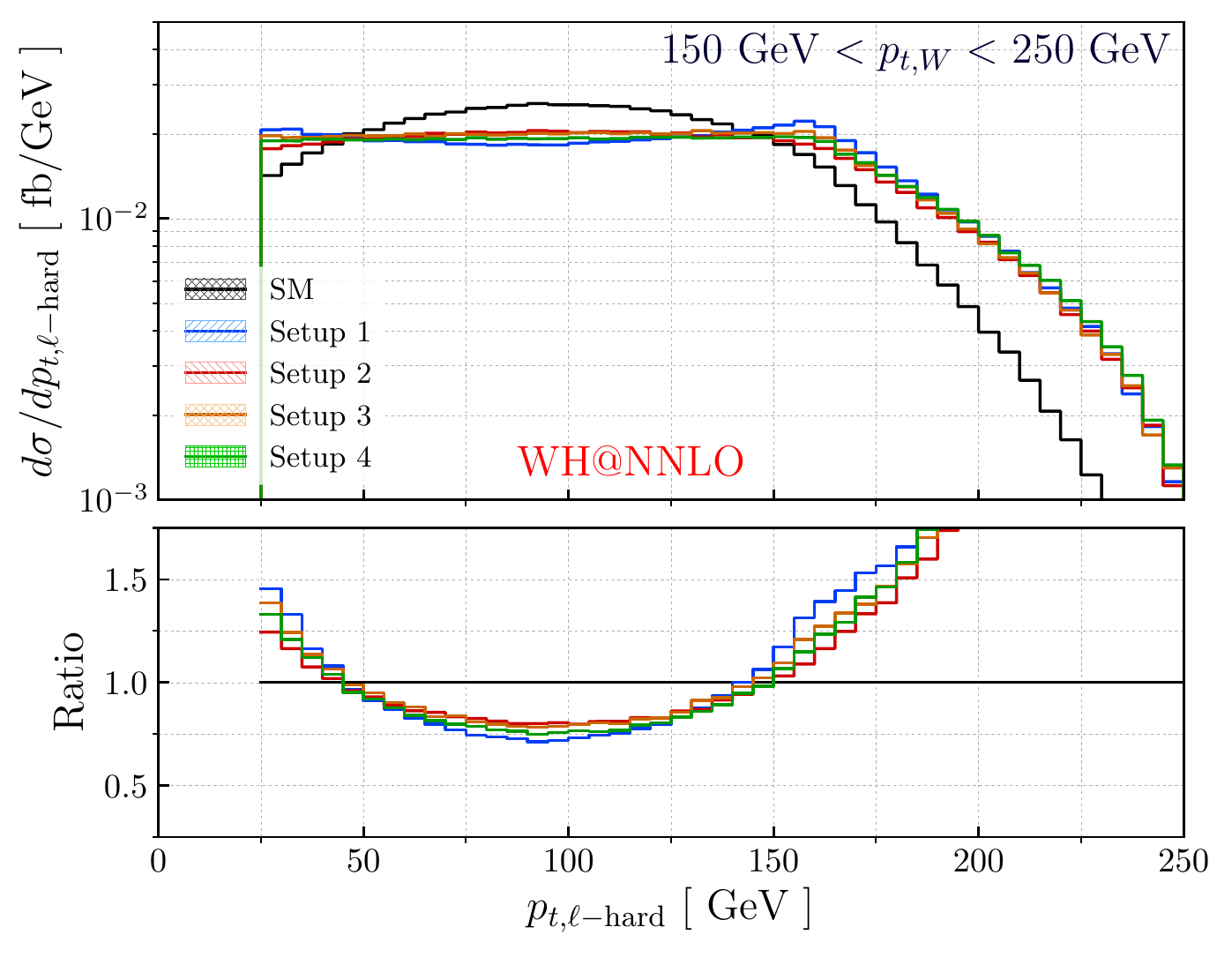}
    \caption{ Kinematic distributions in the process $pp \to
      W^+(e^+\nu)H(b \bar b)$ at the 13 TeV LHC for various SMEFT
      scenarios.  In lower panes ratios of SMEFT to
      SM distributions are shown.  We set the factorization and the
      renormalization scales in the production process to half the
      invariant mass of the $WH$ system.  See text for details.  }
  \label{fig:wh_pth_rec_anom}
\end{figure}

In Fig.~\ref{fig:wh_pth_rec_anom} we show kinematic distributions for the $pp \to W^+H$ process for the four scenarios considered above.
Overall, the situation is similar to what has been already discussed in case of $pp \to ZH$: in peak regions  of all distributions
the different scenarios provide very similar predictions; away from peak regions clear differences are seen in some of them.
These differences, as well as reduced theoretical uncertainties in peak regions, should eventually enable improved studies of the anomalous
couplings in $W^+H$ production.

Before concluding we would like to illustrate the potential impact of the calculations  described in
this paper on bounds on the anomalous couplings that can be obtained from measurements
of fiducial cross sections of $pp \to VH$ processes. 
We
consider a hypothetical measurement of a fiducial $W^+H$ cross section 
 and find the allowed  values  for various combinations of the anomalous couplings.  
We use the same setup as described earlier in this section to define the fiducial $W^+H$ cross section. We assume
that it has been measured and the value 
 $\sigma^{W^+H}_{\rm fid, exp} ={} 3.40(14)~\fb$ was obtained.
We assigned a  four percent   uncertainty to the measured cross section; this corresponds to projections for the high-luminosity  LHC that can be found in
Ref.~\cite{Cepeda:2019klc}.

We  would like to understand how regions of allowed anomalous couplings change when we increase the accuracy of
theoretical computations. 
We note that the fiducial cross section
$\sigma^{WH}_{\rm fid} = f(g_{hww}^{(1)},g_{hww}^{(2)},\tilde{g}_{hww})$ is a polynomial in the couplings.
For this reason, it is enough to sample it for ten different points to determine the full function. 
We then use these results to  check which combinations of anomalous
couplings are compatible with the result of the hypothetical  measurement.
The uncertainty in the experimental cross section  is fixed to four percent and the uncertainties in the theoretical
predictions is determined by varying the scale by a factor of two around the central scale $\mu = M_{WH}/2$. We note that  the
factorization and the renormalization scales are chosen to be equal.

\begin{figure}
  \centering
  \includegraphics[width=0.30\textwidth]{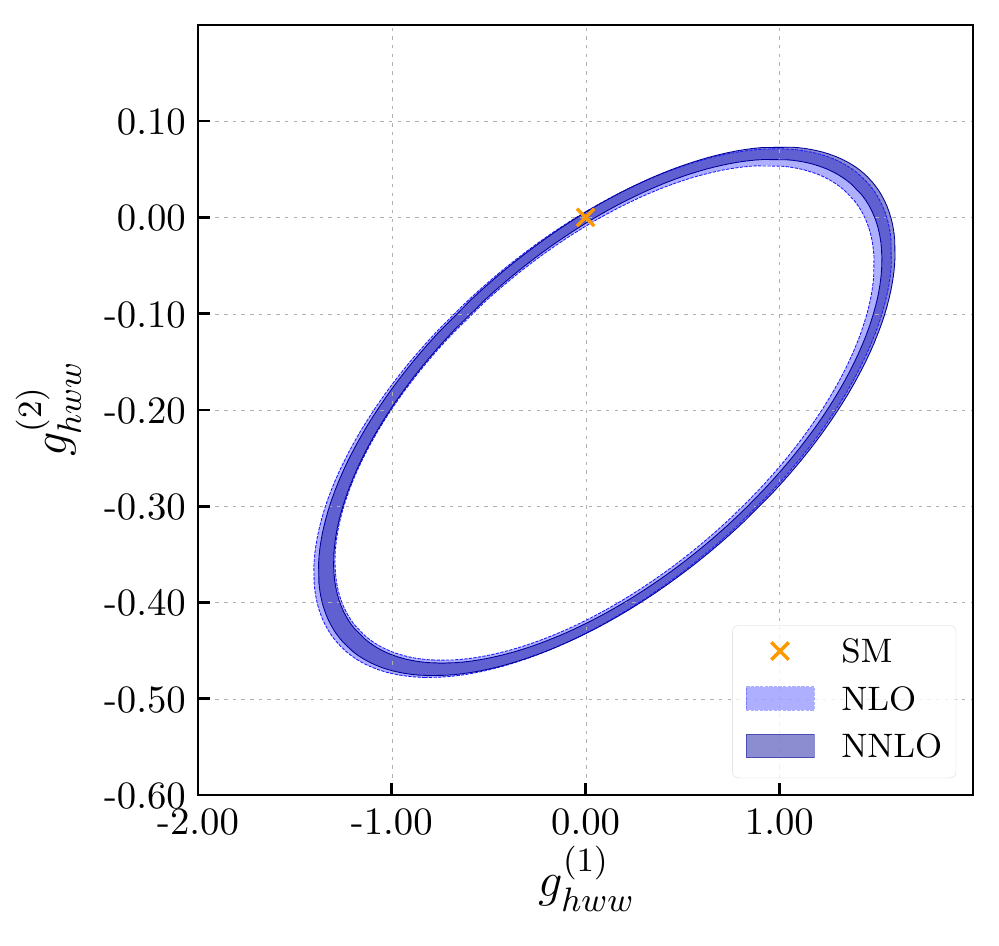}
  \includegraphics[width=0.30\textwidth]{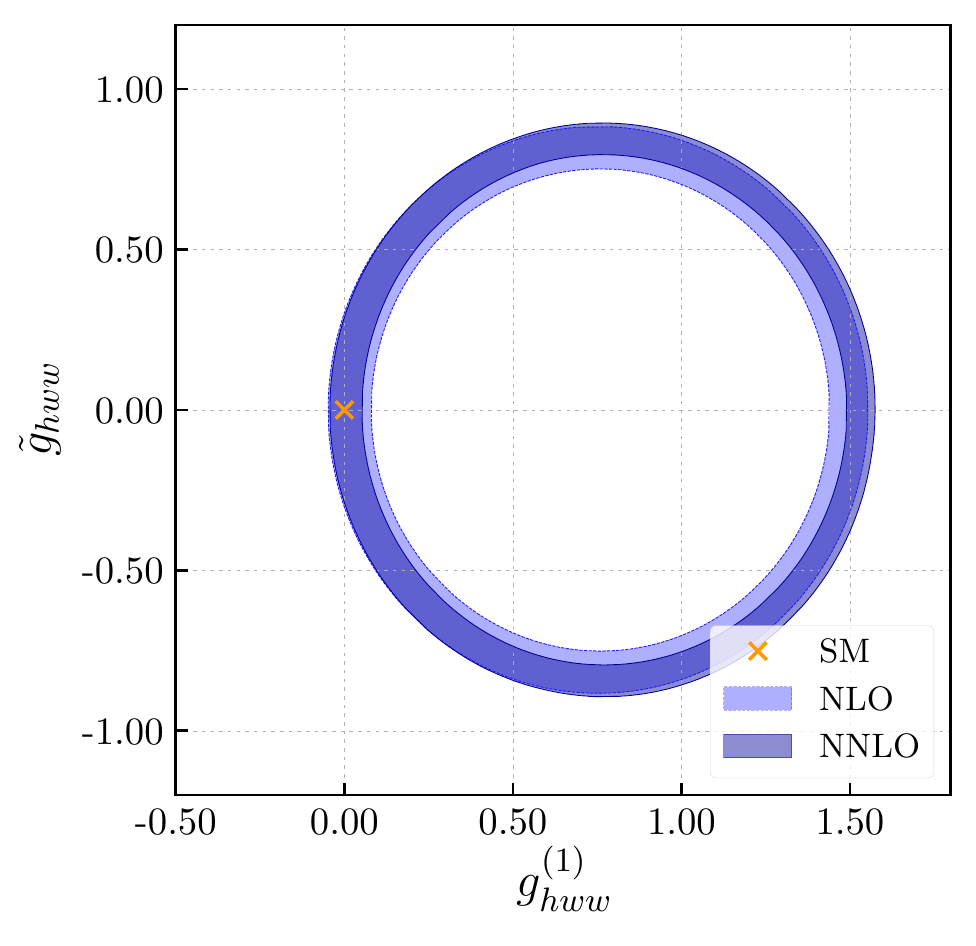}
  \includegraphics[width=0.30\textwidth]{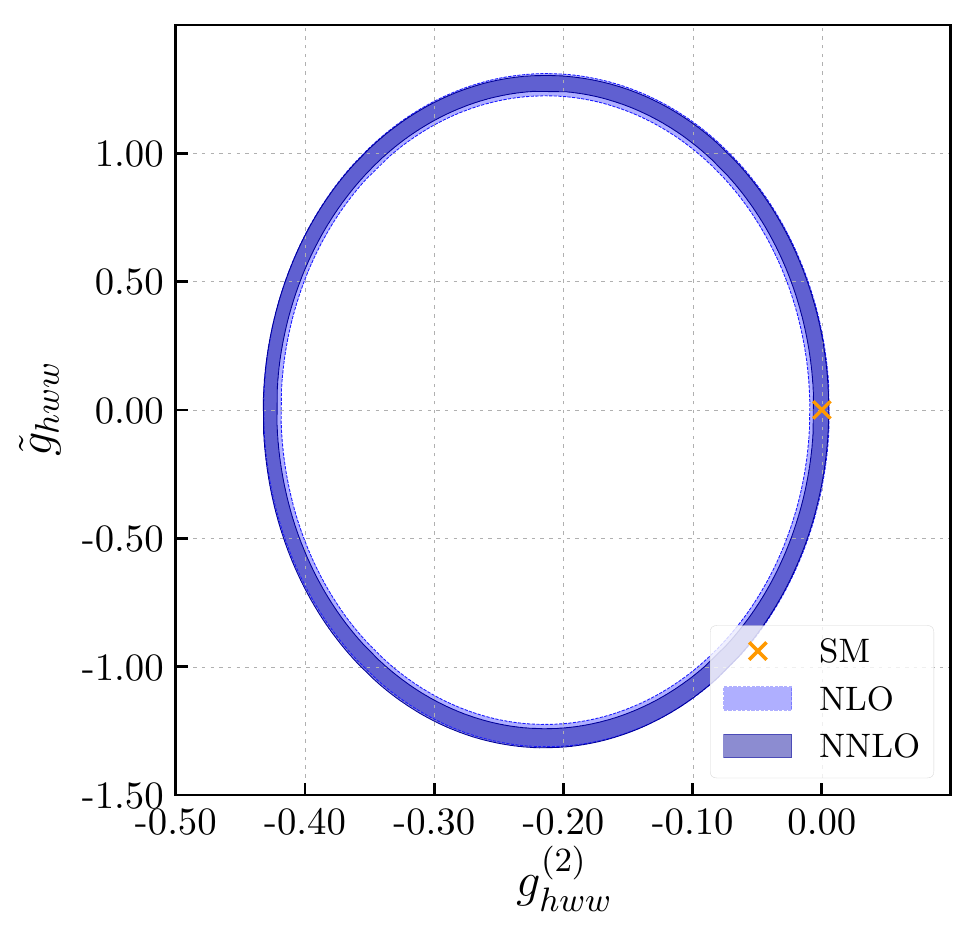}
  \includegraphics[width=0.30\textwidth]{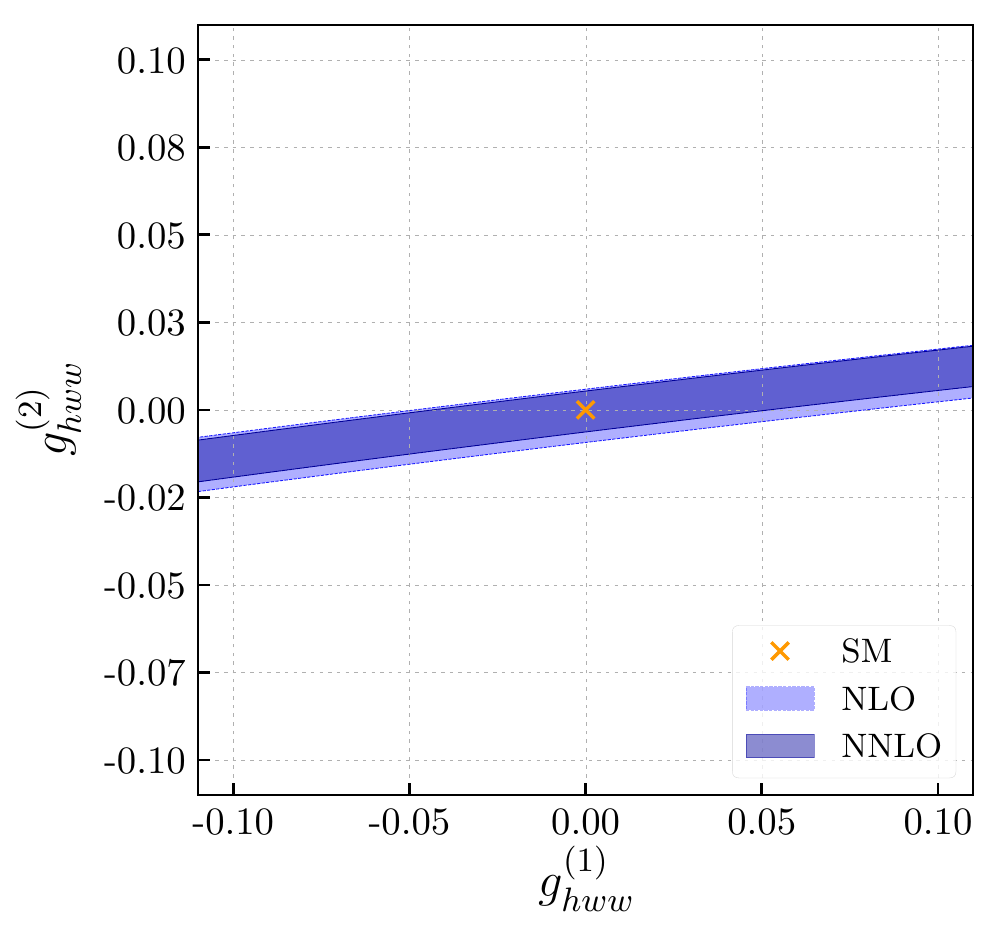}
  \includegraphics[width=0.30\textwidth]{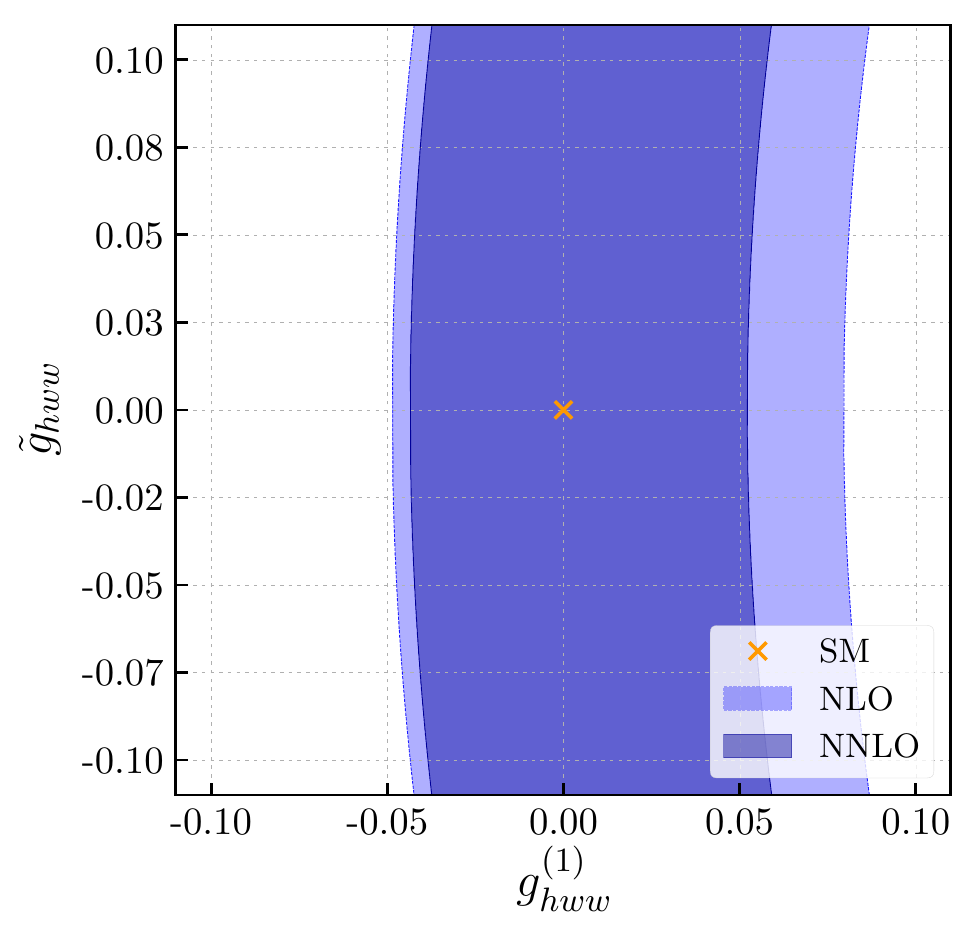}
  \includegraphics[width=0.30\textwidth]{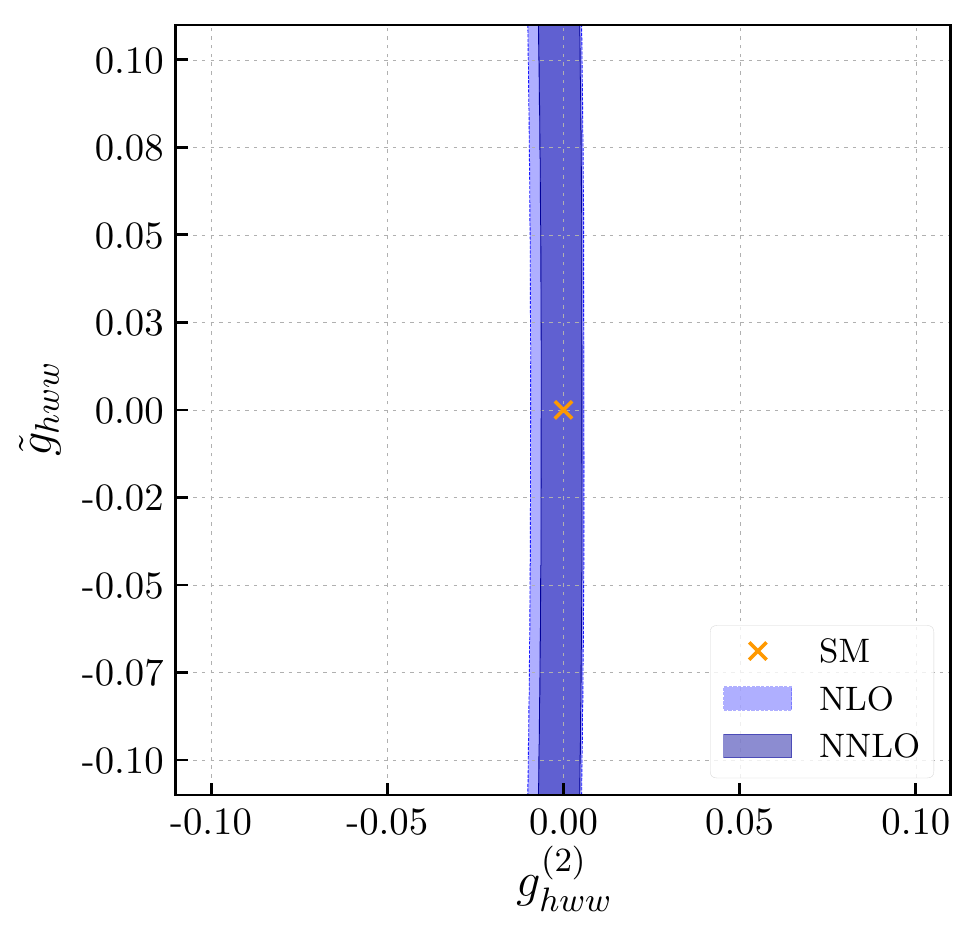}
  \caption{ Examples of contours ($68\%$ confidence level) of allowed
    combinations of anomalous couplings based on a hypothetical
    measurement of the fiducial cross section of $W^+H$ production at the
    13 TeV LHC.
    Color coding describes NLO (lighter blue) and NNLO (darker blue) calculations. The SM result is shown as an orange cross. Upper row: full contours of allowed couplings;
    lower row: contours close to the SM configuration, i.e. for small
    anomalous couplings. A $4\%$ experimental uncertainty was
    assumed. See text for details.}
  \label{fig:wh-contour-HLLHC}
\end{figure}

Our  results are presented  in  Fig.~\ref{fig:wh-contour-HLLHC} where we show
two-dimensional projections of the $g_{hww}^{(i)}$ parameter space.  Shaded areas  mark
couplings  that are compatible with the results of the measurement at the $68\%$  confidence level. 
We note that  NLO QCD corrections change the LO predictions significantly; for this reason, we do not display the latter.
Changes are smaller when moving from NLO to NNLO predictions; 
nevertheless,  we   observe some distortion of shapes of the allowed
region.  This effect is a consequence of the fact that corrections do, in fact, depend
on the anomalous couplings, a feature  that we have already discussed when
talking about Table~\ref{tab:anom-xsec-wh}.

The thickness of the bands representing the allowed regions is only marginally reduced  when NNLO predictions are used
instead of NLO predictions.  This is a consequence of the fact that the experimental uncertainty is fixed at $4$ percent
which is comparable to the scale uncertainty of the theoretical prediction at NLO and is larger than that of the NNLO prediction.

\begin{figure}
  \centering
  \includegraphics[width=0.30\textwidth]{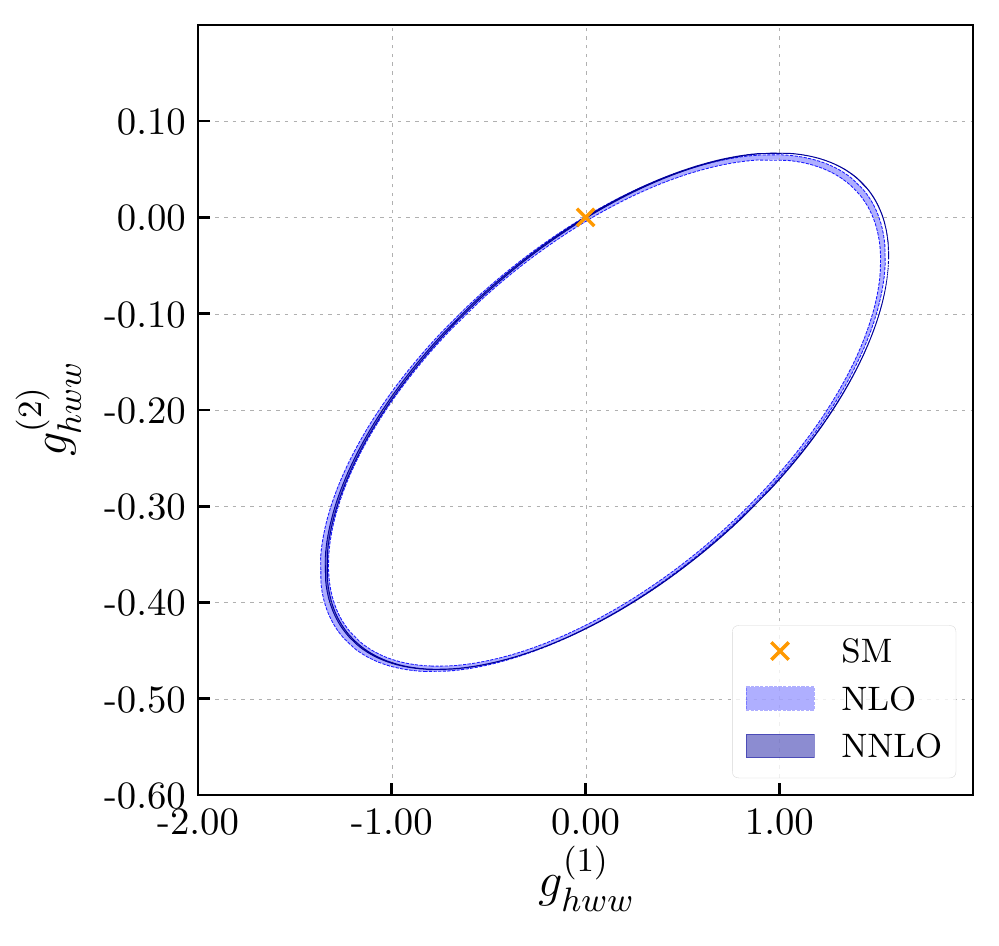}
  \includegraphics[width=0.30\textwidth]{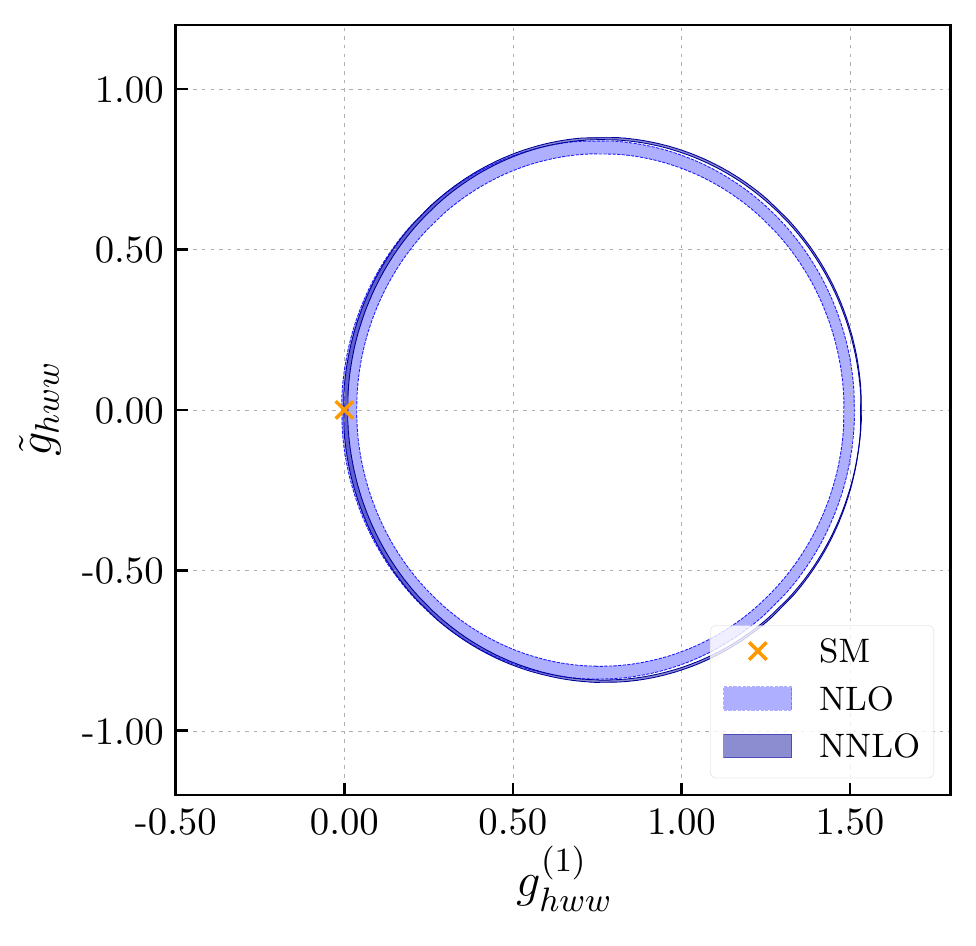}
  \includegraphics[width=0.30\textwidth]{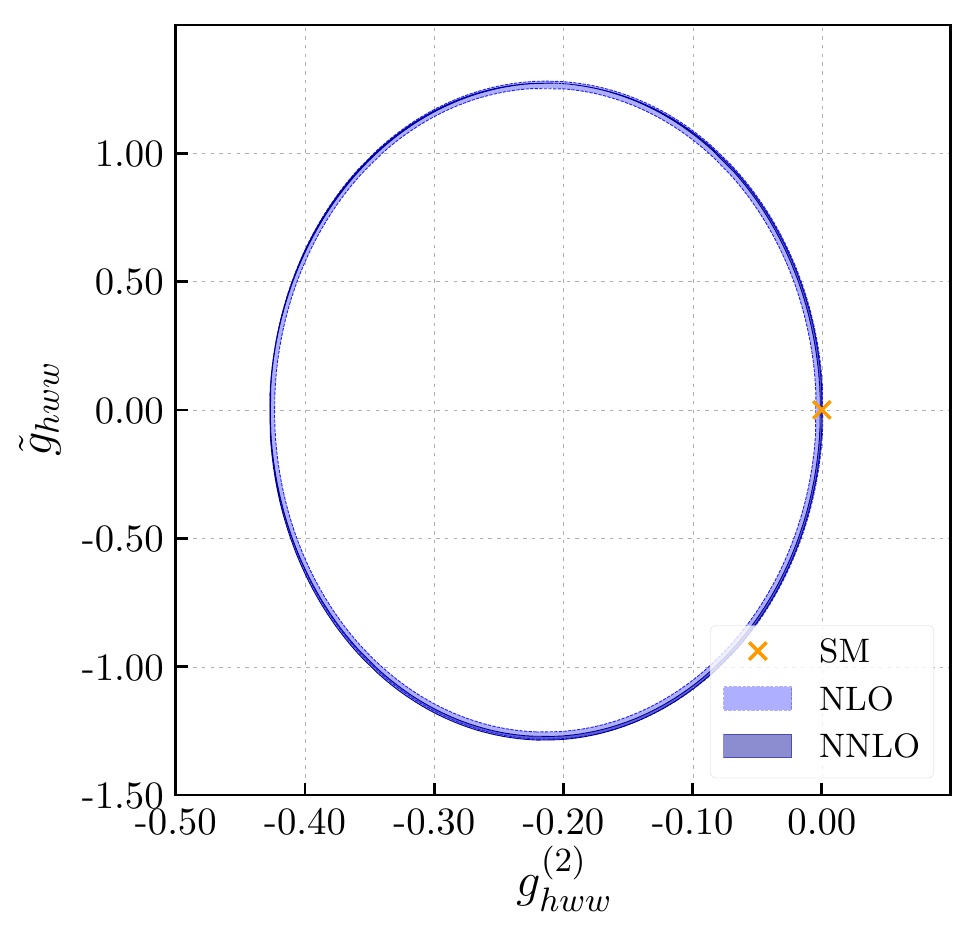}
  \includegraphics[width=0.30\textwidth]{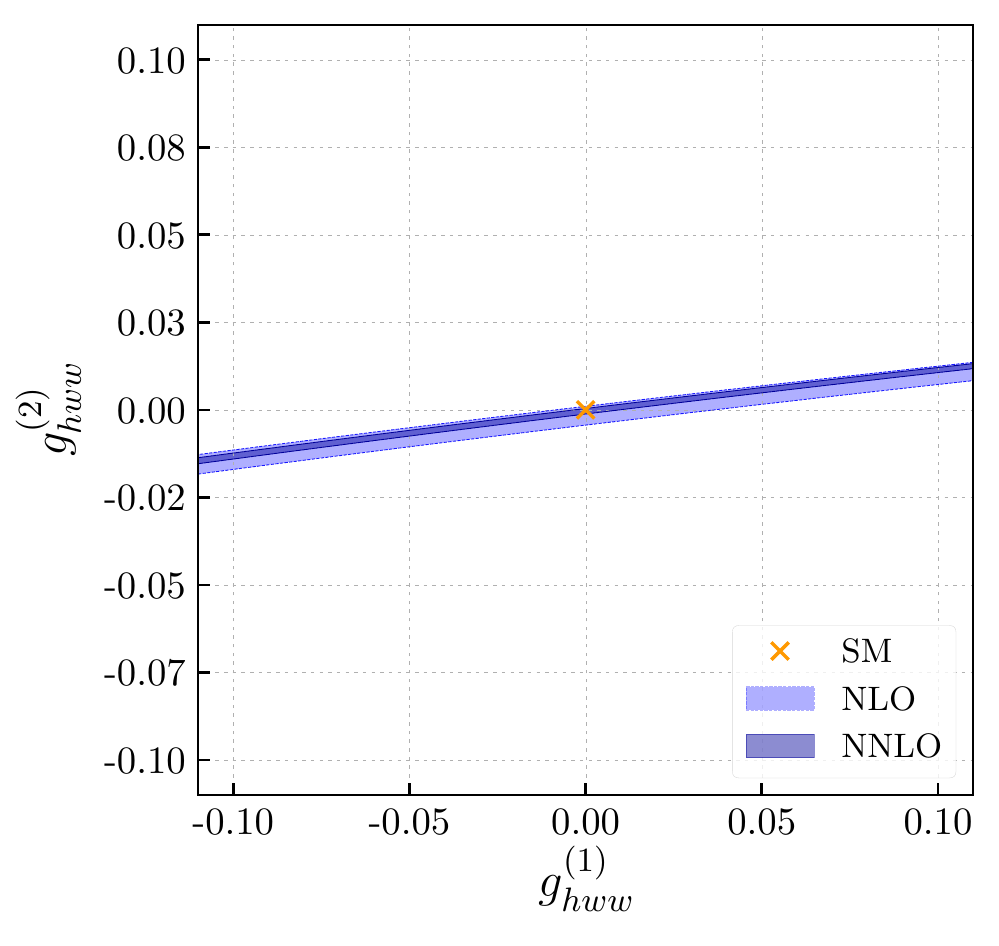}
  \includegraphics[width=0.30\textwidth]{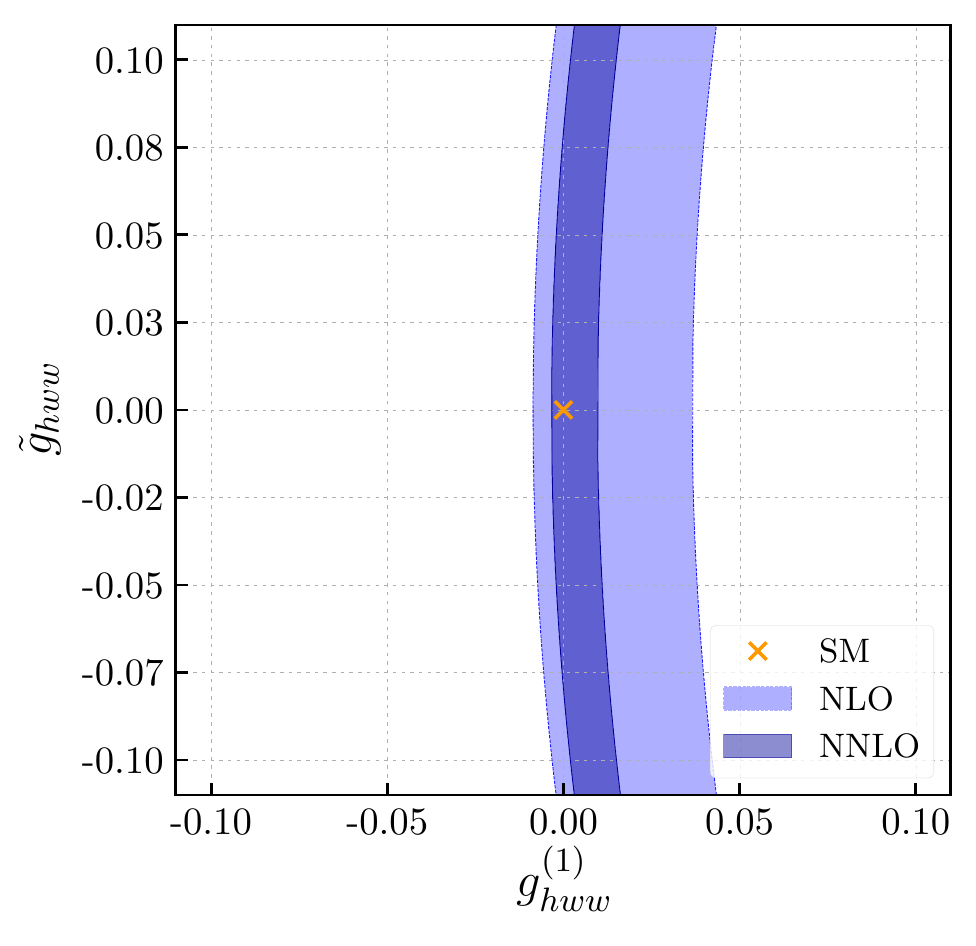}
  \includegraphics[width=0.30\textwidth]{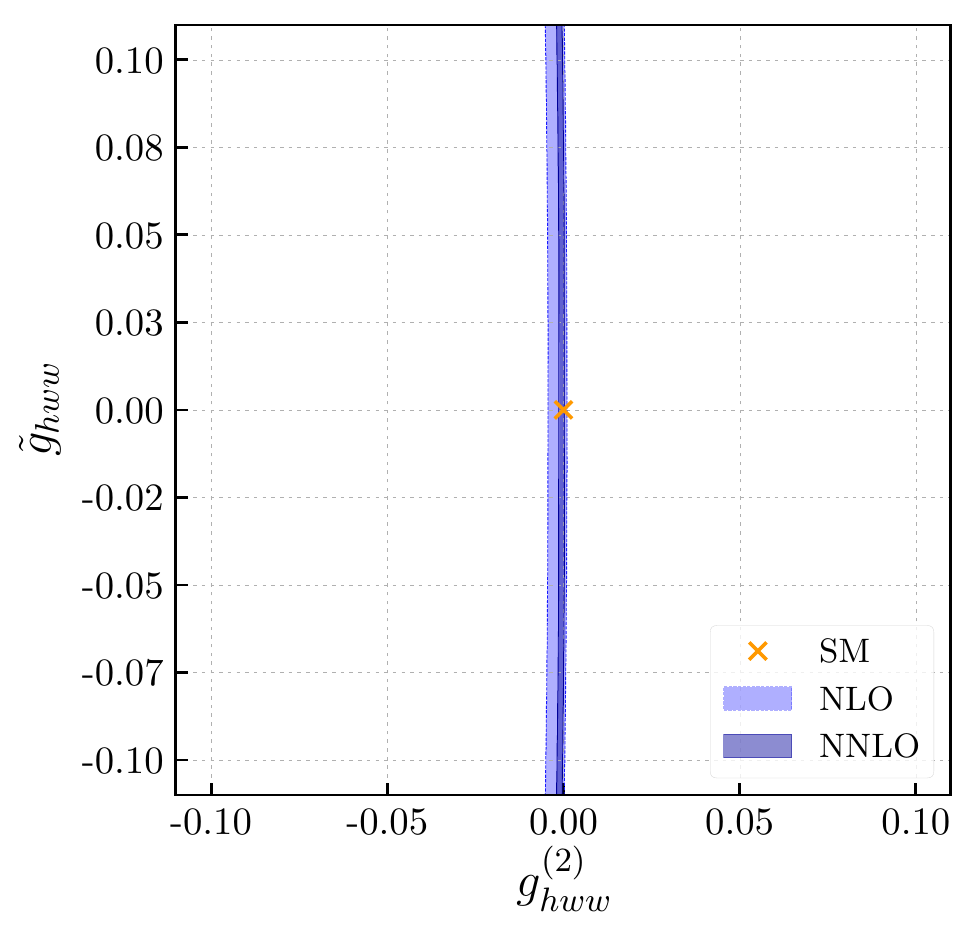}
  \caption{As for Fig.~\ref{fig:wh-contour-HLLHC} but with the experimental uncertainty removed. See text for
       details.
  \label{fig:wh-contour-0pcent}}
\end{figure}

In order to highlight potential  benefits of using NNLO theory predictions, 
we display  similar exclusion limits in
Fig.~\ref{fig:wh-contour-0pcent} but assume that the experimental uncertainty  is significantly reduced. For the sake of argument, we take
it to be zero.
We can now clearly see the thickness of the bands decreasing as we move from NLO to NNLO, as a result of the decreased theoretical error.
Of course, it is unrealistic to assume no experimental uncertainty, but using this assumption does allow us to highlight the benefits of NNLO-accurate theoretical predictions.
We note,  in this regard, that projections in Ref.~\cite{Cepeda:2019klc} are  only estimates  and that it is quite possible 
that the actual results  will outperform these projections. 
If this happens,  we anticipate that
fully-differential NNLO theoretical predictions  will become not only preferable but
perhaps even necessary  for studies of the anomalous couplings in the $pp \to VH$ process.

\section{Conclusions}
\label{sec:summ}

In this paper, we presented computations  of NNLO QCD corrections
to Higgs boson production in association with a $W$ or $Z$ boson.   We included NNLO QCD corrections to $H \to b \bar b$ decays, retaining
full $b$-quark mass dependence. This allowed us to present our results in a setup which is close to the actual experimental analyses.

In addition to NNLO QCD corrections and the effects of the $b$-quark masses, 
our computation also includes  anomalous couplings in the  $VVH$ interaction vertex. We have shown that QCD corrections
to fiducial cross sections depend non-trivially on the anomalous couplings since they  change the relative
importance of various kinematic regions that contribute to  the  fiducial cross sections.
We have argued that the availability of NNLO QCD predictions may allow one to search for the anomalous couplings
in kinematic regions where the EFT framework based on the momentum expansion is  more trustworthy than in the high-energy tails of distributions.
We have also shown  how the NNLO QCD theory predictions may be used to improve exclusion limits for the anomalous couplings; this becomes especially relevant
if experimental uncertainties on fiducial cross sections of  $pp \to VH$  production measured at the high-luminosity LHC reach the few percent
level. 

The computation reported in this paper  describes the most advanced and realistic way  to simulate the associated production process
$pp \to VH$ at the LHC, but further improvements are possible. On the SM side, it is definitely important to refine the calculation
of the $gg \to ZH$ subprocess and to update the contributions to $VH$ production that depend on top loops since many of them are still only known
as expansions in the inverse mass of the top quark.

On the EFT side, one can gradually include contributions of  other dimension-six  
operators, gradually moving towards a full EFT analysis.  Although such extensions of the current computation are not trivial, they are clearly
possible given recent developments in both methods for loop computations and subtraction technology.  We look forward to providing such
refined predictions for $VH$ associated production in the future.

{\bf Acknowledgments:} 
This research is partially supported  by
the Deutsche Forschungsgemeinschaft (DFG, German Research Foundation)
under grant 396021762 - TRR 257. The research of F.C. was partially
supported by the ERC Starting Grant 804394 hipQCD.

\appendix

\bibliographystyle{utphys}
\bibliography{VHassoc}{}

\providecommand{\href}[2]{#2}\begingroup\raggedright\begin{thebibliography}{10}

\bibitem{couplATLCMS}
{\bfseries ATLAS, CMS} Collaboration, G.~Aad {\em et~al.}, ``{Measurements of
  the Higgs boson production and decay rates and constraints on its couplings
  from a combined ATLAS and CMS analysis of the LHC pp collision data at $
  \sqrt{s}=7 $ and 8 TeV},''
  \href{http://dx.doi.org/10.1007/JHEP08(2016)045}{{\em JHEP} {\bfseries 08}
  (2016) 045}, \href{http://arxiv.org/abs/1606.02266}{{\ttfamily
  arXiv:1606.02266 [hep-ex]}}.

\bibitem{Butterworth:2008iy}
J.~M. Butterworth, A.~R. Davison, M.~Rubin, and G.~P. Salam, ``{Jet
  substructure as a new Higgs search channel at the LHC},''
  \href{http://dx.doi.org/10.1103/PhysRevLett.100.242001}{{\em Phys. Rev.
  Lett.} {\bfseries 100} (2008) 242001},
  \href{http://arxiv.org/abs/0802.2470}{{\ttfamily arXiv:0802.2470 [hep-ph]}}.

\bibitem{Aaboud:2017xsd}
{\bfseries ATLAS} Collaboration, M.~Aaboud {\em et~al.}, ``{Evidence for the $
  H\to b\overline{b} $ decay with the ATLAS detector},''
  \href{http://dx.doi.org/10.1007/JHEP12(2017)024}{{\em JHEP} {\bfseries 12}
  (2017) 024}, \href{http://arxiv.org/abs/1708.03299}{{\ttfamily
  arXiv:1708.03299 [hep-ex]}}.

\bibitem{Sirunyan:2017elk}
{\bfseries CMS} Collaboration, A.~M. Sirunyan {\em et~al.}, ``{Evidence for the
  Higgs boson decay to a bottom quark–antiquark pair},''
  \href{http://dx.doi.org/10.1016/j.physletb.2018.02.050}{{\em Phys. Lett.}
  {\bfseries B780} (2018) 501--532},
\href{http://arxiv.org/abs/1709.07497}{{\ttfamily arXiv:1709.07497 [hep-ex]}}.

\bibitem{Aaboud:2018zhk}
{\bfseries ATLAS} Collaboration, M.~Aaboud {\em et~al.}, ``{Observation of $H
  \rightarrow b\bar{b}$ decays and $VH$ production with the ATLAS detector},''
  \href{http://dx.doi.org/10.1016/j.physletb.2018.09.013}{{\em Phys. Lett.}
  {\bfseries B786} (2018) 59--86},
\href{http://arxiv.org/abs/1808.08238}{{\ttfamily arXiv:1808.08238 [hep-ex]}}.

\bibitem{Sirunyan:2018kst}
{\bfseries CMS} Collaboration, A.~M. Sirunyan {\em et~al.}, ``{Observation of
  Higgs boson decay to bottom quarks},''
  \href{http://dx.doi.org/10.1103/PhysRevLett.121.121801}{{\em Phys. Rev.
  Lett.} {\bfseries 121} no.~12, (2018) 121801},
\href{http://arxiv.org/abs/1808.08242}{{\ttfamily arXiv:1808.08242 [hep-ex]}}.

\bibitem{Behring:2020uzq}
A.~Behring, W.~Bizo\'n, F.~Caola, K.~Melnikov, and R.~R\"ontsch, ``{Bottom
  quark mass effects in associated $WH$ production with the $H \to b\bar{b}$
  decay through NNLO QCD},''
  \href{http://dx.doi.org/10.1103/PhysRevD.101.114012}{{\em Phys. Rev. D}
  {\bfseries 101} no.~11, (2020) 114012},
  \href{http://arxiv.org/abs/2003.08321}{{\ttfamily arXiv:2003.08321
  [hep-ph]}}.

\bibitem{Han:1991ia}
T.~Han and S.~Willenbrock, ``{QCD correction to the $p p \to W H$ and $Z H$
  total cross-sections},''
  \href{http://dx.doi.org/10.1016/0370-2693(91)90572-8}{{\em Phys. Lett. B}
  {\bfseries 273} (1991) 167--172}.

\bibitem{Baer:1992vx}
H.~Baer, B.~Bailey, and J.~F. Owens, ``{$\mathcal O (\alpha_s)$ Monte Carlo
  approach to $W$ + Higgs associated production at hadron supercolliders},''
  \href{http://dx.doi.org/10.1103/PhysRevD.47.2730}{{\em Phys. Rev. D}
  {\bfseries 47} (1993) 2730--2734}.

\bibitem{Ohnemus:1992bd}
J.~Ohnemus and W.~J. Stirling, ``{Order $\alpha_s$ corrections to the
  differential cross-section for the $W H$ intermediate mass Higgs signal},''
  \href{http://dx.doi.org/10.1103/PhysRevD.47.2722}{{\em Phys. Rev. D}
  {\bfseries 47} (1993) 2722--2729}.

\bibitem{Mrenna:1997wp}
S.~Mrenna and C.~P. Yuan, ``{Effects of QCD resummation on W+ h and t anti-b
  production at the Tevatron},''
  \href{http://dx.doi.org/10.1016/S0370-2693(97)01303-8}{{\em Phys. Lett. B}
  {\bfseries 416} (1998) 200--207},
  \href{http://arxiv.org/abs/hep-ph/9703224}{{\ttfamily arXiv:hep-ph/9703224}}.

\bibitem{Spira:1997dg}
M.~Spira, ``{QCD effects in Higgs physics},''
  \href{http://dx.doi.org/10.1002/(SICI)1521-3978(199804)46:3<203::AID-PROP203>3.0.CO;2-4}{{\em
  Fortsch. Phys.} {\bfseries 46} (1998) 203--284},
  \href{http://arxiv.org/abs/hep-ph/9705337}{{\ttfamily arXiv:hep-ph/9705337}}.

\bibitem{Djouadi:1999ht}
A.~Djouadi and M.~Spira, ``{SUSY - QCD corrections to Higgs boson production at
  hadron colliders},'' \href{http://dx.doi.org/10.1103/PhysRevD.62.014004}{{\em
  Phys. Rev. D} {\bfseries 62} (2000) 014004},
  \href{http://arxiv.org/abs/hep-ph/9912476}{{\ttfamily arXiv:hep-ph/9912476}}.

\bibitem{Brein:2003wg}
O.~Brein, A.~Djouadi, and R.~Harlander, ``{NNLO QCD corrections to the
  Higgs-strahlung processes at hadron colliders},''
  \href{http://dx.doi.org/10.1016/j.physletb.2003.10.112}{{\em Phys. Lett. B}
  {\bfseries 579} (2004) 149--156},
  \href{http://arxiv.org/abs/hep-ph/0307206}{{\ttfamily arXiv:hep-ph/0307206}}.

\bibitem{Brein:2011vx}
O.~Brein, R.~Harlander, M.~Wiesemann, and T.~Zirke, ``{Top-Quark Mediated
  Effects in Hadronic Higgs-Strahlung},''
  \href{http://dx.doi.org/10.1140/epjc/s10052-012-1868-6}{{\em Eur. Phys. J. C}
  {\bfseries 72} (2012) 1868}, \href{http://arxiv.org/abs/1111.0761}{{\ttfamily
  arXiv:1111.0761 [hep-ph]}}.

\bibitem{Brein:2012ne}
O.~Brein, R.~V. Harlander, and T.~J.~E. Zirke, ``{vh@nnlo - Higgs Strahlung at
  hadron colliders},'' \href{http://dx.doi.org/10.1016/j.cpc.2012.11.002}{{\em
  Comput. Phys. Commun.} {\bfseries 184} (2013) 998--1003},
  \href{http://arxiv.org/abs/1210.5347}{{\ttfamily arXiv:1210.5347 [hep-ph]}}.

\bibitem{Ferrera:2011bk}
G.~Ferrera, M.~Grazzini, and F.~Tramontano, ``{Associated WH production at
  hadron colliders: a fully exclusive QCD calculation at NNLO},''
  \href{http://dx.doi.org/10.1103/PhysRevLett.107.152003}{{\em Phys. Rev.
  Lett.} {\bfseries 107} (2011) 152003},
  \href{http://arxiv.org/abs/1107.1164}{{\ttfamily arXiv:1107.1164 [hep-ph]}}.

\bibitem{Ferrera:2013yga}
G.~Ferrera, M.~Grazzini, and F.~Tramontano, ``{Higher-order QCD effects for
  associated WH production and decay at the LHC},''
  \href{http://dx.doi.org/10.1007/JHEP04(2014)039}{{\em JHEP} {\bfseries 04}
  (2014) 039}, \href{http://arxiv.org/abs/1312.1669}{{\ttfamily arXiv:1312.1669
  [hep-ph]}}.

\bibitem{Ferrera:2014lca}
G.~Ferrera, M.~Grazzini, and F.~Tramontano, ``{Associated ZH production at
  hadron colliders: the fully differential NNLO QCD calculation},''
  \href{http://dx.doi.org/10.1016/j.physletb.2014.11.040}{{\em Phys. Lett. B}
  {\bfseries 740} (2015) 51--55},
  \href{http://arxiv.org/abs/1407.4747}{{\ttfamily arXiv:1407.4747 [hep-ph]}}.

\bibitem{Campbell:2016jau}
J.~M. Campbell, R.~K. Ellis, and C.~Williams, ``{Associated production of a
  Higgs boson at NNLO},'' \href{http://dx.doi.org/10.1007/JHEP06(2016)179}{{\em
  JHEP} {\bfseries 06} (2016) 179},
  \href{http://arxiv.org/abs/1601.00658}{{\ttfamily arXiv:1601.00658
  [hep-ph]}}.

\bibitem{Ferrera:2017zex}
G.~Ferrera, G.~Somogyi, and F.~Tramontano, ``{Associated production of a Higgs
  boson decaying into bottom quarks at the LHC in full NNLO QCD},''
  \href{http://dx.doi.org/10.1016/j.physletb.2018.03.021}{{\em Phys. Lett. B}
  {\bfseries 780} (2018) 346--351},
  \href{http://arxiv.org/abs/1705.10304}{{\ttfamily arXiv:1705.10304
  [hep-ph]}}.

\bibitem{Caola:2017xuq}
F.~Caola, G.~Luisoni, K.~Melnikov, and R.~R\"ontsch, ``{NNLO QCD corrections to
  associated $WH$ production and $H \to b \bar b$ decay},''
  \href{http://dx.doi.org/10.1103/PhysRevD.97.074022}{{\em Phys. Rev. D}
  {\bfseries 97} no.~7, (2018) 074022},
  \href{http://arxiv.org/abs/1712.06954}{{\ttfamily arXiv:1712.06954
  [hep-ph]}}.

\bibitem{Gauld:2019yng}
R.~Gauld, A.~Gehrmann-De~Ridder, E.~W.~N. Glover, A.~Huss, and I.~Majer,
  ``{Associated production of a Higgs boson decaying into bottom quarks and a
  weak vector boson decaying leptonically at NNLO in QCD},''
  \href{http://dx.doi.org/10.1007/JHEP10(2019)002}{{\em JHEP} {\bfseries 10}
  (2019) 002}, \href{http://arxiv.org/abs/1907.05836}{{\ttfamily
  arXiv:1907.05836 [hep-ph]}}.

\bibitem{Ciccolini:2003jy}
M.~L. Ciccolini, S.~Dittmaier, and M.~Kramer, ``{Electroweak radiative
  corrections to associated WH and ZH production at hadron colliders},''
  \href{http://dx.doi.org/10.1103/PhysRevD.68.073003}{{\em Phys. Rev. D}
  {\bfseries 68} (2003) 073003},
  \href{http://arxiv.org/abs/hep-ph/0306234}{{\ttfamily arXiv:hep-ph/0306234}}.

\bibitem{Denner:2011id}
A.~Denner, S.~Dittmaier, S.~Kallweit, and A.~Muck, ``{Electroweak corrections
  to Higgs-strahlung off W/Z bosons at the Tevatron and the LHC with HAWK},''
  \href{http://dx.doi.org/10.1007/JHEP03(2012)075}{{\em JHEP} {\bfseries 03}
  (2012) 075}, \href{http://arxiv.org/abs/1112.5142}{{\ttfamily arXiv:1112.5142
  [hep-ph]}}.

\bibitem{Denner:2014cla}
A.~Denner, S.~Dittmaier, S.~Kallweit, and A.~M\"uck, ``{HAWK 2.0: A Monte Carlo
  program for Higgs production in vector-boson fusion and Higgs strahlung at
  hadron colliders},'' \href{http://dx.doi.org/10.1016/j.cpc.2015.04.021}{{\em
  Comput. Phys. Commun.} {\bfseries 195} (2015) 161--171},
  \href{http://arxiv.org/abs/1412.5390}{{\ttfamily arXiv:1412.5390 [hep-ph]}}.

\bibitem{Gauld:2020ced}
R.~Gauld, A.~Gehrmann-De~Ridder, E.~W.~N. Glover, A.~Huss, and I.~Majer,
  ``{Precise predictions for WH+jet production at the LHC},''
  \href{http://dx.doi.org/10.1016/j.physletb.2021.136335}{{\em Phys. Lett. B}
  {\bfseries 817} (2021) 136335},
  \href{http://arxiv.org/abs/2009.14209}{{\ttfamily arXiv:2009.14209
  [hep-ph]}}.

\bibitem{Kniehl:1990iva}
B.~A. Kniehl, ``{Associated Production of Higgs and Z Bosons From Gluon Fusion
  in Hadron Collisions},''
  \href{http://dx.doi.org/10.1103/PhysRevD.42.2253}{{\em Phys. Rev. D}
  {\bfseries 42} (1990) 2253--2258}.

\bibitem{Dicus:1988yh}
D.~A. Dicus and C.~Kao, ``{Higgs Boson - $Z^0$ Production From Gluon Fusion},''
  \href{http://dx.doi.org/10.1103/PhysRevD.38.1008}{{\em Phys. Rev. D}
  {\bfseries 38} (1988) 1008}. [Erratum: Phys.Rev.D 42, 2412 (1990)].

\bibitem{Altenkamp:2012sx}
L.~Altenkamp, S.~Dittmaier, R.~V. Harlander, H.~Rzehak, and T.~J.~E. Zirke,
  ``{Gluon-induced Higgs-strahlung at next-to-leading order QCD},''
  \href{http://dx.doi.org/10.1007/JHEP02(2013)078}{{\em JHEP} {\bfseries 02}
  (2013) 078}, \href{http://arxiv.org/abs/1211.5015}{{\ttfamily arXiv:1211.5015
  [hep-ph]}}.

\bibitem{Harlander:2014wda}
R.~V. Harlander, A.~Kulesza, V.~Theeuwes, and T.~Zirke, ``{Soft gluon
  resummation for gluon-induced Higgs Strahlung},''
  \href{http://dx.doi.org/10.1007/JHEP11(2014)082}{{\em JHEP} {\bfseries 11}
  (2014) 082}, \href{http://arxiv.org/abs/1410.0217}{{\ttfamily arXiv:1410.0217
  [hep-ph]}}.

\bibitem{Hasselhuhn:2016rqt}
A.~Hasselhuhn, T.~Luthe, and M.~Steinhauser, ``{On top quark mass effects to
  $gg\to ZH$ at NLO},'' \href{http://dx.doi.org/10.1007/JHEP01(2017)073}{{\em
  JHEP} {\bfseries 01} (2017) 073},
  \href{http://arxiv.org/abs/1611.05881}{{\ttfamily arXiv:1611.05881
  [hep-ph]}}.

\bibitem{Davies:2020drs}
J.~Davies, G.~Mishima, and M.~Steinhauser, ``{Virtual corrections to $gg\to ZH$
  in the high-energy and large-$m_t$ limits},''
  \href{http://dx.doi.org/10.1007/JHEP03(2021)034}{{\em JHEP} {\bfseries 03}
  (2021) 034}, \href{http://arxiv.org/abs/2011.12314}{{\ttfamily
  arXiv:2011.12314 [hep-ph]}}.

\bibitem{Chen:2020gae}
L.~Chen, G.~Heinrich, S.~P. Jones, M.~Kerner, J.~Klappert, and J.~Schlenk,
  ``{$ZH$ production in gluon fusion: two-loop amplitudes with full top quark
  mass dependence},'' \href{http://dx.doi.org/10.1007/JHEP03(2021)125}{{\em
  JHEP} {\bfseries 03} (2021) 125},
  \href{http://arxiv.org/abs/2011.12325}{{\ttfamily arXiv:2011.12325
  [hep-ph]}}.

\bibitem{Alasfar:2021ppe}
L.~Alasfar, G.~Degrassi, P.~P. Giardino, R.~Gr\"ober, and M.~Vitti, ``{Virtual
  corrections to $gg\to ZH$ via a transverse momentum expansion},''
  \href{http://dx.doi.org/10.1007/JHEP05(2021)168}{{\em JHEP} {\bfseries 05}
  (2021) 168}, \href{http://arxiv.org/abs/2103.06225}{{\ttfamily
  arXiv:2103.06225 [hep-ph]}}.

\bibitem{Harlander:2018yns}
R.~V. Harlander, J.~Klappert, C.~Pandini, and A.~Papaefstathiou, ``{Exploiting
  the WH/ZH symmetry in the search for New Physics},''
  \href{http://dx.doi.org/10.1140/epjc/s10052-018-6234-x}{{\em Eur. Phys. J. C}
  {\bfseries 78} no.~9, (2018) 760},
  \href{http://arxiv.org/abs/1804.02299}{{\ttfamily arXiv:1804.02299
  [hep-ph]}}.

\bibitem{giuliaWH}
W.~Astill, W.~Bizon, E.~Re, and G.~Zanderighi, ``{NNLOPS accurate associated HW
  production},'' \href{http://dx.doi.org/10.1007/JHEP06(2016)154}{{\em JHEP}
  {\bfseries 06} (2016) 154}, \href{http://arxiv.org/abs/1603.01620}{{\ttfamily
  arXiv:1603.01620 [hep-ph]}}.

\bibitem{Alioli:2019qzz}
S.~Alioli, A.~Broggio, S.~Kallweit, M.~A. Lim, and L.~Rottoli,
  ``{Higgsstrahlung at NNLL'$+$NNLO matched to parton showers in GENEVA},''
  \href{http://dx.doi.org/10.1103/PhysRevD.100.096016}{{\em Phys. Rev. D}
  {\bfseries 100} no.~9, (2019) 096016},
  \href{http://arxiv.org/abs/1909.02026}{{\ttfamily arXiv:1909.02026
  [hep-ph]}}.

\bibitem{giuliaZH}
W.~Astill, W.~Bizo\'n, E.~Re, and G.~Zanderighi, ``{NNLOPS accurate associated
  HZ production with $ H\to b\overline{b} $ decay at NLO},''
  \href{http://dx.doi.org/10.1007/JHEP11(2018)157}{{\em JHEP} {\bfseries 11}
  (2018) 157}, \href{http://arxiv.org/abs/1804.08141}{{\ttfamily
  arXiv:1804.08141 [hep-ph]}}.

\bibitem{Bizon:2019tfo}
W.~Bizo\'n, E.~Re, and G.~Zanderighi, ``{NNLOPS description of the $H \to
  b\overline{b} $ decay with MiNLO},''
  \href{http://dx.doi.org/10.1007/JHEP06(2020)006}{{\em JHEP} {\bfseries 06}
  (2020) 006}, \href{http://arxiv.org/abs/1912.09982}{{\ttfamily
  arXiv:1912.09982 [hep-ph]}}.

\bibitem{Granata:2017iod}
F.~Granata, J.~M. Lindert, C.~Oleari, and S.~Pozzorini, ``{NLO QCD+EW
  predictions for HV and HV +jet production including parton-shower effects},''
  \href{http://dx.doi.org/10.1007/JHEP09(2017)012}{{\em JHEP} {\bfseries 09}
  (2017) 012}, \href{http://arxiv.org/abs/1706.03522}{{\ttfamily
  arXiv:1706.03522 [hep-ph]}}.

\bibitem{Hespel:2015zea}
B.~Hespel, F.~Maltoni, and E.~Vryonidou, ``{Higgs and Z boson associated
  production via gluon fusion in the SM and the 2HDM},''
  \href{http://dx.doi.org/10.1007/JHEP06(2015)065}{{\em JHEP} {\bfseries 06}
  (2015) 065}, \href{http://arxiv.org/abs/1503.01656}{{\ttfamily
  arXiv:1503.01656 [hep-ph]}}.

\bibitem{Gorishnii:1990zu}
S.~G. Gorishnii, A.~L. Kataev, S.~A. Larin, and L.~R. Surguladze, ``{Corrected
  Three Loop {QCD} Correction to the Correlator of the Quark Scalar Currents
  and $\Gamma$ (Tot) ($H^0 \to$ Hadrons)},''
  \href{http://dx.doi.org/10.1142/S0217732390003152}{{\em Mod. Phys. Lett. A}
  {\bfseries 5} (1990) 2703--2712}.

\bibitem{Gorishnii:1991zr}
S.~G. Gorishnii, A.~L. Kataev, S.~A. Larin, and L.~R. Surguladze, ``{Scheme
  dependence of the next to next-to-leading QCD corrections to $\Gamma$(tot)
  $(H^0 \to$ hadrons) and the spurious QCD infrared fixed point},''
  \href{http://dx.doi.org/10.1103/PhysRevD.43.1633}{{\em Phys. Rev. D}
  {\bfseries 43} (1991) 1633--1640}.

\bibitem{Kataev:1993be}
A.~L. Kataev and V.~T. Kim, ``{The Effects of the QCD corrections to $\Gamma
  (H^0 \to b \bar b)$},''
  \href{http://dx.doi.org/10.1142/S0217732394001131}{{\em Mod. Phys. Lett. A}
  {\bfseries 9} (1994) 1309--1326}.

\bibitem{Surguladze:1994gc}
L.~R. Surguladze, ``{Quark mass effects in fermionic decays of the Higgs boson
  in $\mathcal O (\alpha_s^2)$ perturbative QCD},''
  \href{http://dx.doi.org/10.1016/0370-2693(94)01253-9}{{\em Phys. Lett. B}
  {\bfseries 341} (1994) 60--72},
  \href{http://arxiv.org/abs/hep-ph/9405325}{{\ttfamily arXiv:hep-ph/9405325}}.

\bibitem{Larin:1995sq}
S.~A. Larin, T.~van Ritbergen, and J.~A.~M. Vermaseren, ``{The Large top quark
  mass expansion for Higgs boson decays into bottom quarks and into gluons},''
  \href{http://dx.doi.org/10.1016/0370-2693(95)01192-S}{{\em Phys. Lett. B}
  {\bfseries 362} (1995) 134--140},
  \href{http://arxiv.org/abs/hep-ph/9506465}{{\ttfamily arXiv:hep-ph/9506465}}.

\bibitem{Chetyrkin:1995pd}
K.~G. Chetyrkin and A.~Kwiatkowski, ``{Second order QCD corrections to scalar
  and pseudoscalar Higgs decays into massive bottom quarks},''
  \href{http://dx.doi.org/10.1016/0550-3213(95)00616-8}{{\em Nucl. Phys. B}
  {\bfseries 461} (1996) 3--18},
  \href{http://arxiv.org/abs/hep-ph/9505358}{{\ttfamily arXiv:hep-ph/9505358}}.

\bibitem{Chetyrkin:1996sr}
K.~G. Chetyrkin, ``{Correlator of the quark scalar currents and $\Gamma$(tot)
  ($H \to$ hadrons) at $\mathcal O (\alpha_s^3)$ in pQCD},''
  \href{http://dx.doi.org/10.1016/S0370-2693(96)01368-8}{{\em Phys. Lett. B}
  {\bfseries 390} (1997) 309--317},
  \href{http://arxiv.org/abs/hep-ph/9608318}{{\ttfamily arXiv:hep-ph/9608318}}.

\bibitem{Baikov:2005rw}
P.~A. Baikov, K.~G. Chetyrkin, and J.~H. Kuhn, ``{Scalar correlator at
  $\mathcal O(\alpha_s^4)$, Higgs decay into b-quarks and bounds on the light
  quark masses},'' \href{http://dx.doi.org/10.1103/PhysRevLett.96.012003}{{\em
  Phys. Rev. Lett.} {\bfseries 96} (2006) 012003},
  \href{http://arxiv.org/abs/hep-ph/0511063}{{\ttfamily arXiv:hep-ph/0511063}}.

\bibitem{Davies:2017xsp}
J.~Davies, M.~Steinhauser, and D.~Wellmann, ``{Completing the hadronic Higgs
  boson decay at order $\alpha_s^4$},''
  \href{http://dx.doi.org/10.1016/j.nuclphysb.2017.04.012}{{\em Nucl. Phys. B}
  {\bfseries 920} (2017) 20--31},
  \href{http://arxiv.org/abs/1703.02988}{{\ttfamily arXiv:1703.02988
  [hep-ph]}}.

\bibitem{Fleischer:1980ub}
J.~Fleischer and F.~Jegerlehner, ``{Radiative Corrections to Higgs Decays in
  the Extended Weinberg-Salam Model},''
  \href{http://dx.doi.org/10.1103/PhysRevD.23.2001}{{\em Phys. Rev. D}
  {\bfseries 23} (1981) 2001--2026}.

\bibitem{Bardin:1990zj}
D.~Y. Bardin, B.~M. Vilensky, and P.~K. Khristova, ``{Calculation of the Higgs
  boson decay width into fermion pairs},'' {\em Sov. J. Nucl. Phys.} {\bfseries
  53} (1991) 152--158.

\bibitem{Dabelstein:1991ky}
A.~Dabelstein and W.~Hollik, ``{Electroweak corrections to the fermionic decay
  width of the standard Higgs boson},''
  \href{http://dx.doi.org/10.1007/BF01625912}{{\em Z. Phys. C} {\bfseries 53}
  (1992) 507--516}.

\bibitem{Kniehl:1991ze}
B.~A. Kniehl, ``{Radiative corrections for $H \to$ f anti-f ($\gamma$) in the
  standard model},'' \href{http://dx.doi.org/10.1016/0550-3213(92)90065-J}{{\em
  Nucl. Phys. B} {\bfseries 376} (1992) 3--28}.

\bibitem{Denner:2011mq}
A.~Denner, S.~Heinemeyer, I.~Puljak, D.~Rebuzzi, and M.~Spira, ``{Standard
  Model Higgs-Boson Branching Ratios with Uncertainties},''
  \href{http://dx.doi.org/10.1140/epjc/s10052-011-1753-8}{{\em Eur. Phys. J. C}
  {\bfseries 71} (2011) 1753}, \href{http://arxiv.org/abs/1107.5909}{{\ttfamily
  arXiv:1107.5909 [hep-ph]}}.

\bibitem{Spira:2016ztx}
M.~Spira, ``{Higgs Boson Production and Decay at Hadron Colliders},''
  \href{http://dx.doi.org/10.1016/j.ppnp.2017.04.001}{{\em Prog. Part. Nucl.
  Phys.} {\bfseries 95} (2017) 98--159},
  \href{http://arxiv.org/abs/1612.07651}{{\ttfamily arXiv:1612.07651
  [hep-ph]}}.

\bibitem{Anastasiou:2011qx}
C.~Anastasiou, F.~Herzog, and A.~Lazopoulos, ``{The fully differential decay
  rate of a Higgs boson to bottom-quarks at NNLO in QCD},''
  \href{http://dx.doi.org/10.1007/JHEP03(2012)035}{{\em JHEP} {\bfseries 03}
  (2012) 035}, \href{http://arxiv.org/abs/1110.2368}{{\ttfamily arXiv:1110.2368
  [hep-ph]}}.

\bibitem{DelDuca:2015zqa}
V.~Del~Duca, C.~Duhr, G.~Somogyi, F.~Tramontano, and Z.~Tr\'ocs\'anyi, ``{Higgs
  boson decay into b-quarks at NNLO accuracy},''
  \href{http://dx.doi.org/10.1007/JHEP04(2015)036}{{\em JHEP} {\bfseries 04}
  (2015) 036}, \href{http://arxiv.org/abs/1501.07226}{{\ttfamily
  arXiv:1501.07226 [hep-ph]}}.

\bibitem{Caola:2019pfz}
F.~Caola, K.~Melnikov, and R.~R\"ontsch, ``{Analytic results for decays of
  color singlets to $gg$ and $q \bar q$ final states at NNLO QCD with the
  nested soft-collinear subtraction scheme},''
  \href{http://dx.doi.org/10.1140/epjc/s10052-019-7505-x}{{\em Eur. Phys. J. C}
  {\bfseries 79} no.~12, (2019) 1013},
  \href{http://arxiv.org/abs/1907.05398}{{\ttfamily arXiv:1907.05398
  [hep-ph]}}.

\bibitem{Mondini:2019vub}
R.~Mondini and C.~Williams, ``{$ H\to b\overline{b}j $ at
  next-to-next-to-leading order accuracy},''
  \href{http://dx.doi.org/10.1007/JHEP06(2019)120}{{\em JHEP} {\bfseries 06}
  (2019) 120}, \href{http://arxiv.org/abs/1904.08961}{{\ttfamily
  arXiv:1904.08961 [hep-ph]}}.

\bibitem{Mondini:2019gid}
R.~Mondini, M.~Schiavi, and C.~Williams, ``{N$^{3}$LO predictions for the decay
  of the Higgs boson to bottom quarks},''
  \href{http://dx.doi.org/10.1007/JHEP06(2019)079}{{\em JHEP} {\bfseries 06}
  (2019) 079}, \href{http://arxiv.org/abs/1904.08960}{{\ttfamily
  arXiv:1904.08960 [hep-ph]}}.

\bibitem{Bernreuther:2018ynm}
W.~Bernreuther, L.~Chen, and Z.-G. Si, ``{Differential decay rates of CP-even
  and CP-odd Higgs bosons to top and bottom quarks at NNLO QCD},''
  \href{http://dx.doi.org/10.1007/JHEP07(2018)159}{{\em JHEP} {\bfseries 07}
  (2018) 159}, \href{http://arxiv.org/abs/1805.06658}{{\ttfamily
  arXiv:1805.06658 [hep-ph]}}.

\bibitem{Behring:2019oci}
A.~Behring and W.~Bizo\'n, ``{Higgs decay into massive b-quarks at NNLO QCD in
  the nested soft-collinear subtraction scheme},''
  \href{http://dx.doi.org/10.1007/JHEP01(2020)189}{{\em JHEP} {\bfseries 01}
  (2020) 189}, \href{http://arxiv.org/abs/1911.11524}{{\ttfamily
  arXiv:1911.11524 [hep-ph]}}.

\bibitem{Somogyi:2020mmk}
G.~Somogyi and F.~Tramontano, ``{Fully exclusive heavy quark-antiquark pair
  production from a colourless initial state at NNLO in QCD},''
  \href{http://dx.doi.org/10.1007/JHEP11(2020)142}{{\em JHEP} {\bfseries 11}
  (2020) 142}, \href{http://arxiv.org/abs/2007.15015}{{\ttfamily
  arXiv:2007.15015 [hep-ph]}}.

\bibitem{Primo:2018zby}
A.~Primo, G.~Sasso, G.~Somogyi, and F.~Tramontano, ``{Exact Top Yukawa
  corrections to Higgs boson decay into bottom quarks},''
  \href{http://dx.doi.org/10.1103/PhysRevD.99.054013}{{\em Phys. Rev. D}
  {\bfseries 99} no.~5, (2019) 054013},
  \href{http://arxiv.org/abs/1812.07811}{{\ttfamily arXiv:1812.07811
  [hep-ph]}}.

\bibitem{Mondini:2020uyy}
R.~Mondini, U.~Schubert, and C.~Williams, ``{Top-induced contributions to
  $H\rightarrow b\bar{b}$ and $H\rightarrow c\bar{c}$ at
  $\mathcal{O}(\alpha_s^3)$},''
  \href{http://dx.doi.org/10.1007/JHEP12(2020)058}{{\em JHEP} {\bfseries 12}
  (2020) 058}, \href{http://arxiv.org/abs/2006.03563}{{\ttfamily
  arXiv:2006.03563 [hep-ph]}}.

\bibitem{Brivio:2017vri}
I.~Brivio and M.~Trott, ``{The Standard Model as an Effective Field Theory},''
  \href{http://dx.doi.org/10.1016/j.physrep.2018.11.002}{{\em Phys. Rept.}
  {\bfseries 793} (2019) 1--98},
  \href{http://arxiv.org/abs/1706.08945}{{\ttfamily arXiv:1706.08945
  [hep-ph]}}.

\bibitem{Mimasu:2015nqa}
K.~Mimasu, V.~Sanz, and C.~Williams, ``{Higher Order QCD predictions for
  Associated Higgs production with anomalous couplings to gauge bosons},''
  \href{http://dx.doi.org/10.1007/JHEP08(2016)039}{{\em JHEP} {\bfseries 08}
  (2016) 039}, \href{http://arxiv.org/abs/1512.02572}{{\ttfamily
  arXiv:1512.02572 [hep-ph]}}.

\bibitem{Caola:2017dug}
F.~Caola, K.~Melnikov, and R.~R\"ontsch, ``{Nested soft-collinear subtractions
  in NNLO QCD computations},''
  \href{http://dx.doi.org/10.1140/epjc/s10052-017-4774-0}{{\em Eur. Phys. J. C}
  {\bfseries 77} no.~4, (2017) 248},
  \href{http://arxiv.org/abs/1702.01352}{{\ttfamily arXiv:1702.01352
  [hep-ph]}}.

\bibitem{Caola:2019nzf}
F.~Caola, K.~Melnikov, and R.~R\"ontsch, ``{Analytic results for color-singlet
  production at NNLO QCD with the nested soft-collinear subtraction scheme},''
  \href{http://dx.doi.org/10.1140/epjc/s10052-019-6880-7}{{\em Eur. Phys. J. C}
  {\bfseries 79} no.~5, (2019) 386},
  \href{http://arxiv.org/abs/1902.02081}{{\ttfamily arXiv:1902.02081
  [hep-ph]}}.

\bibitem{hxswg}
{\bfseries LHC Higgs Cross Section Working Group} Collaboration, D.~de~Florian
  {\em et~al.}, ``{Handbook of LHC Higgs Cross Sections: 4. Deciphering the
  Nature of the Higgs Sector},''
  \href{http://arxiv.org/abs/1610.07922}{{\ttfamily arXiv:1610.07922
  [hep-ph]}}.

\bibitem{Chetyrkin:2000yt}
K.~G. Chetyrkin, J.~H. K{\"u}hn, and M.~Steinhauser, ``{RunDec: A Mathematica
  package for running and decoupling of the strong coupling and quark
  masses},'' \href{http://dx.doi.org/10.1016/S0010-4655(00)00155-7}{{\em
  Comput. Phys. Commun.} {\bfseries 133} (2000) 43--65},
\href{http://arxiv.org/abs/hep-ph/0004189}{{\ttfamily arXiv:hep-ph/0004189
  [hep-ph]}}.

\bibitem{Herren:2017osy}
F.~Herren and M.~Steinhauser, ``{Version 3 of RunDec and CRunDec},''
  \href{http://dx.doi.org/10.1016/j.cpc.2017.11.014}{{\em Comput. Phys.
  Commun.} {\bfseries 224} (2018) 333--345},
\href{http://arxiv.org/abs/1703.03751}{{\ttfamily arXiv:1703.03751 [hep-ph]}}.

\bibitem{Cacciari:2008gp}
M.~Cacciari, G.~P. Salam, and G.~Soyez, ``{The anti-$k_t$ jet clustering
  algorithm},'' \href{http://dx.doi.org/10.1088/1126-6708/2008/04/063}{{\em
  JHEP} {\bfseries 04} (2008) 063},
  \href{http://arxiv.org/abs/0802.1189}{{\ttfamily arXiv:0802.1189 [hep-ph]}}.

\bibitem{Cacciari:2011ma}
M.~Cacciari, G.~P. Salam, and G.~Soyez, ``{FastJet User Manual},''
  \href{http://dx.doi.org/10.1140/epjc/s10052-012-1896-2}{{\em Eur. Phys. J. C}
  {\bfseries 72} (2012) 1896}, \href{http://arxiv.org/abs/1111.6097}{{\ttfamily
  arXiv:1111.6097 [hep-ph]}}.

\bibitem{Englert:2013vua}
C.~Englert, M.~McCullough, and M.~Spannowsky, ``{Gluon-initiated associated
  production boosts Higgs physics},''
  \href{http://dx.doi.org/10.1103/PhysRevD.89.013013}{{\em Phys. Rev. D}
  {\bfseries 89} no.~1, (2014) 013013},
  \href{http://arxiv.org/abs/1310.4828}{{\ttfamily arXiv:1310.4828 [hep-ph]}}.

\bibitem{Farina:2012xp}
M.~Farina, C.~Grojean, F.~Maltoni, E.~Salvioni, and A.~Thamm, ``{Lifting
  degeneracies in Higgs couplings using single top production in association
  with a Higgs boson},'' \href{http://dx.doi.org/10.1007/JHEP05(2013)022}{{\em
  JHEP} {\bfseries 05} (2013) 022},
  \href{http://arxiv.org/abs/1211.3736}{{\ttfamily arXiv:1211.3736 [hep-ph]}}.

\bibitem{Sirunyan:2018lzm}
{\bfseries CMS} Collaboration, A.~M. Sirunyan {\em et~al.}, ``{Search for
  associated production of a Higgs boson and a single top quark in
  proton-proton collisions at $\sqrt{s} = 13$ TeV},''
  \href{http://dx.doi.org/10.1103/PhysRevD.99.092005}{{\em Phys. Rev. D}
  {\bfseries 99} no.~9, (2019) 092005},
  \href{http://arxiv.org/abs/1811.09696}{{\ttfamily arXiv:1811.09696
  [hep-ex]}}.

\bibitem{Aad:2020jym}
{\bfseries ATLAS} Collaboration, G.~Aad {\em et~al.}, ``{Measurements of $WH$
  and $ZH$ production in the $H \rightarrow b\bar{b}$ decay channel in $pp$
  collisions at 13 TeV with the ATLAS detector},''
  \href{http://dx.doi.org/10.1140/epjc/s10052-020-08677-2}{{\em Eur. Phys. J.
  C} {\bfseries 81} no.~2, (2021) 178},
  \href{http://arxiv.org/abs/2007.02873}{{\ttfamily arXiv:2007.02873
  [hep-ex]}}.

\bibitem{Cepeda:2019klc}
M.~Cepeda {\em et~al.}, ``{Report from Working Group 2}: {Higgs Physics at the
  HL-LHC and HE-LHC},''
  \href{http://dx.doi.org/10.23731/CYRM-2019-007.221}{{\em CERN Yellow Rep.
  Monogr.} {\bfseries 7} (2019) 221--584},
  \href{http://arxiv.org/abs/1902.00134}{{\ttfamily arXiv:1902.00134
  [hep-ph]}}.

\end{thebibliography}\endgroup

\end{document}